\documentclass[10pt, conference]{IEEEtran}
\IEEEoverridecommandlockouts

\usepackage{cite}
\usepackage{amsmath,amssymb,amsfonts}
\usepackage{graphicx}
\usepackage{textcomp}
\usepackage{xcolor}
\usepackage{xspace}
\usepackage{enumitem}
\def\BibTeX{{\rm B\kern-.05em{\sc i\kern-.025em b}\kern-.08em
    T\kern-.1667em\lower.7ex\hbox{E}\kern-.125emX}}

\usepackage{xcolor,colortbl}
\usepackage{float}
\usepackage{graphicx}
\usepackage{subfigure}
\usepackage{multirow}
\usepackage{xcolor}
\usepackage{colortbl}
\usepackage{fancybox}

\usepackage{setspace}
\usepackage{enumitem}
\usepackage{color}
\usepackage{diagbox}
\usepackage{soul}
\usepackage{algpseudocode}
\usepackage{framed}
\usepackage{amsmath}
\usepackage{pifont}
\usepackage{balance}
\usepackage[normalem]{ulem}
\usepackage{url}
\usepackage[T1]{fontenc}
\usepackage[utf8]{inputenc}
\usepackage[font=small,labelfont=bf,tableposition=top]{caption}
\usepackage{booktabs}
\usepackage{threeparttable}
\usepackage{fontawesome5}
\usepackage[many]{tcolorbox}

\usepackage[ruled,vlined,linesnumbered]{algorithm2e}

\definecolor{grey}{rgb}{0.9,0.9,0.9}
\definecolor{lightgreen}{HTML}{bae4b3}
\definecolor{lightgrey}{HTML}{f0f0f0}
\definecolor{mygreen}{HTML}{31a354}
\definecolor{mygray}{HTML}{666666}

\newcommand*{\mycode}{\fontfamily{lmtt}\selectfont}

\makeatletter
\newcommand{\linebreakand}{%
  \end{@IEEEauthorhalign}
  \hfill\mbox{}\par
  \mbox{}\hfill\begin{@IEEEauthorhalign}
}
\makeatother

\renewcommand{\baselinestretch}{.97}

 \usepackage[colorlinks]{hyperref}   
 
\definecolor{mygreen}{HTML}{31a354}

\def\BibTeX{{\rm B\kern-.05em{\sc i\kern-.025em b}\kern-.08em
    T\kern-.1667em\lower.7ex\hbox{E}\kern-.125emX}}

\usepackage{xcolor,colortbl}
\usepackage{float}
\usepackage{graphicx}
\usepackage{subfigure}
\usepackage{multirow}
\usepackage{xcolor}
\usepackage{colortbl}
\usepackage{fancybox}
\hypersetup{colorlinks,linkcolor={blue},citecolor={blue},urlcolor={red}} 
\hypersetup{colorlinks=true, urlcolor=blue}
\hypersetup{colorlinks,linkcolor={blue},citecolor={blue},urlcolor={red}} 
\hypersetup{colorlinks=true, urlcolor=blue}

\AtBeginDocument{
  \hypersetup{pdfborder={0 0 1}, urlcolor=.}  
}

\begin{document}

\title{MCTS-Refined CoT: High-Quality Fine-Tuning Data for LLM-Based Repository Issue Resolution}
\author{
\centering
\IEEEauthorblockN{Yibo Wang}
\IEEEauthorblockA{
    Northeastern University\\
    Shenyang, China \\
    yibowangcz@outlook.com
}
\and
\IEEEauthorblockN{Zhihao Peng}
\IEEEauthorblockA{
    Northeastern University\\
    Shenyang, China \\
    2471378@stu.neu.edu.cn
}

\and
\IEEEauthorblockN{Ying Wang\IEEEauthorrefmark{1}}
\IEEEauthorblockA{
    Northeastern University\\
    Shenyang, China \\
    wangying@swc.neu.edu.cn
}

\linebreakand
\IEEEauthorblockN{Zhao Wei\IEEEauthorrefmark{1}}
\IEEEauthorblockA{
    Tencent\\
    Beijing, China \\
    zachwei@tencent.com
}
\and
\IEEEauthorblockN{Hai Yu}
\IEEEauthorblockA{
    Northeastern University\\
    Shenyang, China \\
    yuhai@mail.neu.edu.cn
}
\and
\IEEEauthorblockN{Zhiliang Zhu}
\IEEEauthorblockA{
    Northeastern University\\
    Shenyang, China \\
    ZHUZhiLiang\_NEU@163.com
}
\thanks{* Ying Wang and Zhao Wei are the corresponding authors.}
}

\maketitle
\vspace{-4mm}
\begin{abstract}

LLMs demonstrate strong performance in automated software engineering, particularly for code generation and issue resolution. While proprietary models like \emph{GPT-4o} achieve high benchmarks scores on \emph{SWE-bench}, their API dependence, cost, and privacy concerns limit adoption. Open-source alternatives offer transparency but underperform in complex tasks, especially sub-100B parameter models. Although quality Chain-of-Thought (CoT) data can enhance reasoning, current methods face two critical flaws: (1) weak rejection sampling reduces data quality, and (2) inadequate step validation causes error accumulation. These limitations lead to flawed reasoning chains that impair LLMs' ability to learn reliable issue resolution.

The paper proposes \textsc{MCTS-Refine}, an enhanced Monte Carlo Tree Search (MCTS)-based algorithm that dynamically validates and optimizes intermediate reasoning steps through a rigorous rejection sampling strategy, generating high-quality CoT data to improve LLM performance in issue resolution tasks. Key innovations include: (1) augmenting MCTS with a reflection mechanism that corrects errors via rejection sampling and refinement, (2) decomposing issue resolution into three subtasks—\emph{File Localization}, \emph{Fault Localization}, and \emph{Patch Generation}—each with clear ground-truth criteria, and (3) enforcing a strict sampling protocol where intermediate outputs must exactly match verified developer patches, ensuring correctness across reasoning paths.

Experiments on \emph{SWE-bench Lite} and \emph{SWE-bench Verified} demonstrate that LLMs fine-tuned with our CoT dataset achieve substantial improvements over baselines. Notably, \emph{Qwen2.5-72B-Instruct} achieves \textcolor{black}{28.3}\%(\emph{Lite}) and \textcolor{black}{35.0}\%(\emph{Verified}) resolution rates, surpassing SOTA baseline \emph{SWE-Fixer-Qwen-\textbf{72B}} with the same parameter scale, which only reached \textcolor{black}{24.7}\%(\emph{Lite}) and \textcolor{black}{32.8}\%(\emph{Verified}).
Given precise issue locations as input, our fine-tuned \emph{Qwen2.5-72B-Instruct} model achieves an impressive issue resolution rate of 43.8\%(\emph{Verified}), comparable to the performance of \emph{Deepseek-v3}.
We open-source our \textsc{MCTS-Refine} framework, CoT dataset, and fine-tuned models to advance research in AI-driven software engineering.

\end{abstract}

\section{Introduction}

In recent years, large language models (LLMs) have demonstrated remarkable capabilities in tackling complex tasks, particularly achieving significant progress in code generation and issue resolution within the field of software engineering~\cite{fan2023large,hou2024large,wang2024software}. By comprehending and reasoning about large codebases, LLMs can efficiently perform tasks such as code completion~\cite{liu2024Deepseek,guo2024Deepseek,Roziere2023codellama}, issue patching~\cite{SWE-Bench}, and code refactoring~\cite{Cordeiro2024An}. 
State-of-the-art automated software development tools primarily rely on closed-source models (e.g., \emph{GPT-4o}~\cite{GPT-4o} and \emph{Claude-3.5-Sonnet}~\cite{Claude-3.5-Sonnet}), which achieve strong performance on repository-level issue resolution benchmarks like \emph{SWE-bench Lite} and \emph{SWE-bench Verified}~\cite{SWE-Bench}.
However, the dependence on external APIs, high invocation costs, and potential privacy concerns associated with closed-source models limit their widespread adoption in real-world scenarios. In contrast, open-source LLMs have emerged as a crucial alternative for advancing automated software development due to their flexibility, transparency, and controllability. However, open-source LLMs (<100B parameters) demonstrate significantly inferior performance in issue resolution tasks compared to their closed-source counterparts, with this performance gap substantially constraining their practical utility~\cite{SWE-GPT}.

Recent advances demonstrate that generating high-quality Chain-of-Thought (CoT) data and employing them for fine-tuning represents an effective methodology for enhancing open-source LLMs' complex task performance~\cite{yag2024chain,li2025structured}. Training LLMs with CoT data containing explicit reasoning steps enables superior contextual comprehension and stepwise solution derivation. 
Research~\cite{CoT,qu2024recursive,tang2025realcritic} indicates that \textbf{providing LLMs with correct reasoning steps—even minimally critical information—can substantially enhance performance, whereas erroneous reasoning steps significantly impair their capabilities. This demonstrates that high-quality CoT data is essential for advancing the reasoning performance of LLMs.}

Existing approaches~\cite{SWE-GPT, SWE-Gym, Sorft, SWE-RL} for repository-level issue resolution typically employ single-turn Q\&A prompting to generate CoT data. This involves synthesizing complete reasoning paths by providing issue descriptions and repository context, followed by performance enhancement through supervised fine-tuning (SFT)~\cite{zhang2023sft} or reinforcement learning (RL)~
\cite{guo2025deepseekr1}. However, these methods still face two key limitations:

\begin{itemize}[leftmargin=*, topsep=1pt, itemsep=1pt]
	
\item \textit{\textbf{Inadequate Rejection Sampling Mechanisms for CoT Generation:}} Existing approaches to CoT data generation frequently employ permissive rejection sampling criteria - typically using validation metrics like \emph{Jaccard Similarity} against developer-approved issue patch records (\emph{ground truth}). While this strategy enhances generation efficiency, it risks compromising dataset quality by allowing defective reasoning chains to persist (see \S~\ref{Limitation 1}).

\item \textit{\textbf{Lack of Systematic Validation for Intermediate Reasoning Steps in CoT:}} Existing approaches typically employ ``\emph{Single-turn Q\&A prompting}''—providing LLMs with issue descriptions and repository context to generate issue resolution CoT data—while only validating the final output and neglecting intermediate reasoning steps. When logical errors or omissions occur in these intermediate steps, they propagate through subsequent reasoning, ultimately compromising the quality of CoT data (see \S~\ref{Limitation 2}).

\end{itemize}

The recent successful applications of Monte Carlo Tree Search (MCTS) in multiple domains offer a promising solution to the aforementioned challenges~\cite{zhang2024rest,chen2025towards,guan2025rstar}. By exploring multiple reasoning paths, MCTS can effectively enhance the efficiency of solving complex tasks. Although MCTS has demonstrated significant potential in machine translation and data reasoning tasks, its application in generating high-quality CoT data and improving LLMs' performance in repository-level issue resolution remains underexplored.

To address these challenges, we propose \textsc{MCTS-Refine}, an enhanced \emph{Monte Carlo Tree Search}-based algorithm. Our approach dynamically validates and optimizes intermediate reasoning steps through a rigorous CoT sampling strategy, effectively constructing high-quality CoT data \textsc{MCoT} to significantly improve LLM performance in issue resolution tasks. The key innovations of \textsc{MCTS-Refine} include:

\begin{itemize}[leftmargin=*, topsep=1pt, itemsep=1pt]
	
\item \textit{\textbf{MCTS with Reflective Mechanism:}} We enhance standard MCTS (\emph{\textbf{Selection}}, \emph{\textbf{Expansion}}, \emph{\textbf{Simulation}}, and \emph{\textbf{Backpropagation}}) with a reflection mechanism that validates intermediate reasoning steps against ground truth. By integrating rejection sampling and refinement phases, our approach dynamically corrects errors and optimizes reasoning paths, significantly improving CoT data quality.

\item \textit{\textbf{Subtask Decomposition for Issue Resolution:}} We decompose the issue resolution process into three distinct subtasks: \emph{\textbf{File Localization}}, \emph{\textbf{Fault Localization}}, and \emph{\textbf{Patch Generation}}. This hierarchical decomposition not only reduces task complexity but also establishes clear ground truth criteria for each subtask. By integrating with the \textsc{MCTS-Refine} algorithm, our approach generates high-quality CoT data at each subtask level, ensuring that every intermediate output effectively contributes to the final solution.

\item \textit{\textbf{Rigorous Chain-of-Thought Sampling Protocol:}}
For each of the three subtasks, we implement a strict rejection sampling mechanism based on their respective ground truth criteria. Specifically: (1) \emph{\textbf{File Localization}} requires reasoning paths to produce file paths that exactly match those in developer-verified patches. (2) \emph{\textbf{Fault Localization}} demands precise identification of classes/methods that completely align with ground truth. (3) \emph{\textbf{Patch Generation}} enforces generation of modified code that is identical to the actual developer patches.
This rigorous sampling protocol enables \textsc{MCTS-Refine} to guarantee both correctness and consistency throughout the reasoning paths.

\end{itemize}

To validate the effectiveness of \textsc{MCTS-Refine}, we conducted supervised fine-tuning on open-source LLMs: \emph{Qwen2.5-Coder-7B-Instruct}, \emph{Qwen2.5-Coder-32B-Instruct}, and \emph{Qwen2.5-72B-Instruct}~\cite{qwenmodels} using the \textsc{MCTS-Refine}-generated CoT data, followed by experimental evaluation on the \emph{SWE-bench} benchmark. Our evaluation results demonstrate significant performance improvements across \emph{\textbf{File Localization}}, \emph{\textbf{Fault Localization}}, and \emph{\textbf{Patch Generation}} subtasks for our fine-tuned LLMs. Our key findings are: 

(1) Fine-tuned \emph{Qwen2.5-\textbf{72B}-Instruct} achieved resolution rates of \textcolor{blue}{28.3}\% and \textcolor{blue}{35.0}\% on \emph{SWE-Bench-Lite} and \emph{SWE-Bench-Verified}, respectively, surpassing SOTA baseline \emph{SWE-Fixer-Qwen-\textbf{72B}} with the same parameter scale, which only reached \textcolor{blue}{24.7}\% and \textcolor{blue}{32.8}\%.



(2) Fine-tuned \emph{Qwen2.5-Coder-\textbf{32B}-Instruct} achieved resolution rates of \textcolor{blue}{25.7}\% and \textcolor{blue}{32.4}\% on \emph{SWE-Bench-Lite} and \emph{SWE-Bench-Verified}, respectively, surpassing the SOTA baseline \emph{SoRFT-Qwen-\textbf{32B}} with the same parameter scale, which scored \textcolor{blue}{24.0}\% and \textcolor{blue}{30.8}\%.



(3) Fine-tuned \emph{Qwen2.5-Coder-\textbf{7B}-Instruct} achieved resolution rates of \textcolor{blue}{16.3}\% and \textcolor{blue}{22.6}\% on \emph{SWE-Bench-Lite} and \emph{SWE-Bench-Verified}, respectively, outperforming the SOTA baseline \emph{SoRFT-Qwen-\textbf{7B}} with the same parameter scale, which scored \textcolor{blue}{14.0}\% and \textcolor{blue}{21.4}\%.

(4) Given precise issue locations as input, our fine-tuned \emph{Qwen2.5-72B-Instruct} model achieves an impressive issue resolution rate of \textcolor{blue}{43.8}\%, comparable to the performance of \emph{Deepseek-v3}.



This paper makes the following contributions:

\begin{itemize}[leftmargin=*, topsep=1pt, itemsep=1pt]
	
\item \textit{\textbf{MCTS-Refine Framework:}} We present \textsc{MCTS-Refine}, a framework for automated generation of high-quality CoT data for issue resolution. Our method integrates MCTS with a reflection mechanism to dynamically optimize reasoning steps, while decomposing the task into three structured subtasks with rigorous ground-truth alignment, ensuring superior data quality.

\item \textit{\textbf{Substantial Performance Improvements:}} Fine-tuning LLMs with our \textsc{MCTS-Refine}-generated dataset yields significant enhancements in reasoning performance across all subtasks. Experimental results demonstrate state-of-the-art performance on both \emph{SWE-bench Lite} and \emph{SWE-bench Verified} benchmarks, establishing a new paradigm for applying LLMs to real-world software engineering challenges.

\item \textit{\textbf{Open-sourced Artifacts:}} We open-source both the \textsc{MCoT} dataset generated by \textsc{MCTS-Refine} and our fine-tuned LLMs on our website (\textbf{\emph{https://mcts-refine.github.io/}}), to facilitate further research.

\end{itemize}
\section{Limitations of Existing Work}
\label{Limitations}

To address complex reasoning challenges in repository-level issue resolution, recent studies~\cite{SWE-GPT, SWE-Gym, SWE-Fixer, Sorft, SWE-RL} employ a two-stage approach: (1) Leveraging state-of-the-art LLMs (e.g., \emph{GPT-4o}~\cite{GPT-4o}, \emph{Claude-3.5-Sonnet}~\cite{Claude-3.5-Sonnet}) to generate phased CoT data for issue localization and patch generation; (2) Fine-tuning parameter-efficient open-source foundation models (e.g., \emph{LLaMA3-7B}~\cite{llama3}, \emph{Qwen2.5-7B}~\cite{qwenmodels}) with the synthesized CoT data to enhance task-specific reasoning performance.
Table~\ref{Intro} provides a comparison of existing work~\cite{SWE-GPT, SWE-Gym, SWE-Fixer, Sorft, SWE-RL}, highlighting their distinctive features.
Although these approaches have achieved notable progress in repository-level issue resolution, they exhibit two critical limitations in CoT data synthesis: \textcircled{1} \textbf{\emph{inadequate rejection sampling mechanisms for CoT generation}}, and \textcircled{2} \textbf{\emph{lack of systematic validation for intermediate reasoning steps in CoT}}.
\textbf{When exposed to CoT datasets contaminated by defective reasoning steps during fine-tuning, LLMs cannot acquire the essential logical schemata for reliable issue resolution.}

\begin{table*}[]
\scriptsize
\centering
\setlength\tabcolsep{5pt}
\def\arraystretch{1.1}
\vspace{-8mm}
\caption{A comparison of existing works' distinctive features}
\vspace{-2mm}
 \bgroup
\begin{tabular}{|c|cc|c|cc|}
\hline
\rowcolor[HTML]{000000} 
\cellcolor[HTML]{000000}{\color[HTML]{FFFFFF} }                                             & \multicolumn{2}{c|}{\cellcolor[HTML]{000000}{\color[HTML]{FFFFFF} \textit{\textbf{Sub-Task CoT}}}}                                                       & \cellcolor[HTML]{000000}{\color[HTML]{FFFFFF} }                                                                                                        & \multicolumn{2}{c|}{\cellcolor[HTML]{000000}{\color[HTML]{FFFFFF} \textit{\textbf{Rejection Sampling for CoT}}}}                                         \\ \cline{2-3} \cline{5-6} 
\rowcolor[HTML]{9B9B9B} 
\multirow{-2}{*}{\cellcolor[HTML]{000000}{\color[HTML]{FFFFFF} \textit{\textbf{Approach}}}} & \multicolumn{1}{c|}{\cellcolor[HTML]{9B9B9B}{\color[HTML]{000000} \textit{\textbf{Localization}}}} & {\color[HTML]{000000} \textit{\textbf{Generation}}} & \multirow{-2}{*}{\cellcolor[HTML]{FFFFFF}{\textit{\textbf{\color[HTML]{000000}\begin{tabular}[c]{@{}c@{}}Validation for \\ CoT Steps\end{tabular}}}}} & \multicolumn{1}{c|}{\cellcolor[HTML]{9B9B9B}{\color[HTML]{000000} \textit{\textbf{Localization}}}} & {\color[HTML]{000000} \textit{\textbf{Generation}}} \\ \hline \hline
\textit{\textbf{Lingma-SWE GPT}~\cite{SWE-GPT}}                                                            & \multicolumn{1}{c|}{\textcolor{green!80!black}{\textbf{\checkmark}}}                               & \textcolor{green!80!black}{\textbf{\checkmark}}     & \textcolor{red}{\ding{55}}                                                                                                                             & \multicolumn{1}{c|}{\textit{Jaccard Similarity}}                                                   & \textit{CodeBLEU}                                   \\ \hline
\rowcolor[HTML]{C0C0C0} 
\textit{\textbf{\textsc{Sorft}}~\cite{Sorft}}                                                                     & \multicolumn{1}{c|}{\cellcolor[HTML]{C0C0C0}\textcolor{green!80!black}{\textbf{\checkmark}}}       & \textcolor{green!80!black}{\textbf{\checkmark}}     & \textcolor{red}{\ding{55}}                                                                                                                             & \multicolumn{1}{c|}{\cellcolor[HTML]{C0C0C0}\textit{Overlap with the locations of gold patch}}         & \textit{Overlap with the content of gold patch}         \\ \hline
\textit{\textbf{SWE-RL}~\cite{SWE-RL}}                                                                    & \multicolumn{1}{c|}{\textcolor{green!80!black}{\textbf{\checkmark}}}                               & \textcolor{green!80!black}{\textbf{\checkmark}}     & \textcolor{red}{\ding{55}}                                                                                                                             & \multicolumn{1}{c|}{\textit{Include the file of gold patch}}                                       & \textit{Correct code diff format}                   \\ \hline
\rowcolor[HTML]{C0C0C0} 
\textit{\textbf{SWE-Gym}~\cite{SWE-Gym}}                                                                   & \multicolumn{1}{c|}{\cellcolor[HTML]{C0C0C0}\textcolor{red}{\ding{55}}}                            & \textcolor{green!80!black}{\textbf{\checkmark}}     & \textcolor{red}{\ding{55}}                                                                                                                             & \multicolumn{1}{c|}{\cellcolor[HTML]{C0C0C0}\textcolor{red}{\ding{55}}}                            & Unit Test                                           \\ \hline
\textit{\textbf{SWE-Fixer}~\cite{SWE-Fixer}}                                                                 & \multicolumn{1}{c|}{\textcolor{red}{\ding{55}}}                                                    & \textcolor{green!80!black}{\textbf{\checkmark}}     & \textcolor{red}{\ding{55}}                                                                                                                             & \multicolumn{1}{c|}{\textcolor{red}{\ding{55}}}                                                    & \textcolor{red}{\ding{55}}                          \\ \hline
\end{tabular}
\egroup
\vspace{-4mm}
\label{Intro}
\end{table*}

\begin{figure*}[]
	\centering
\includegraphics[width=0.92\textwidth]{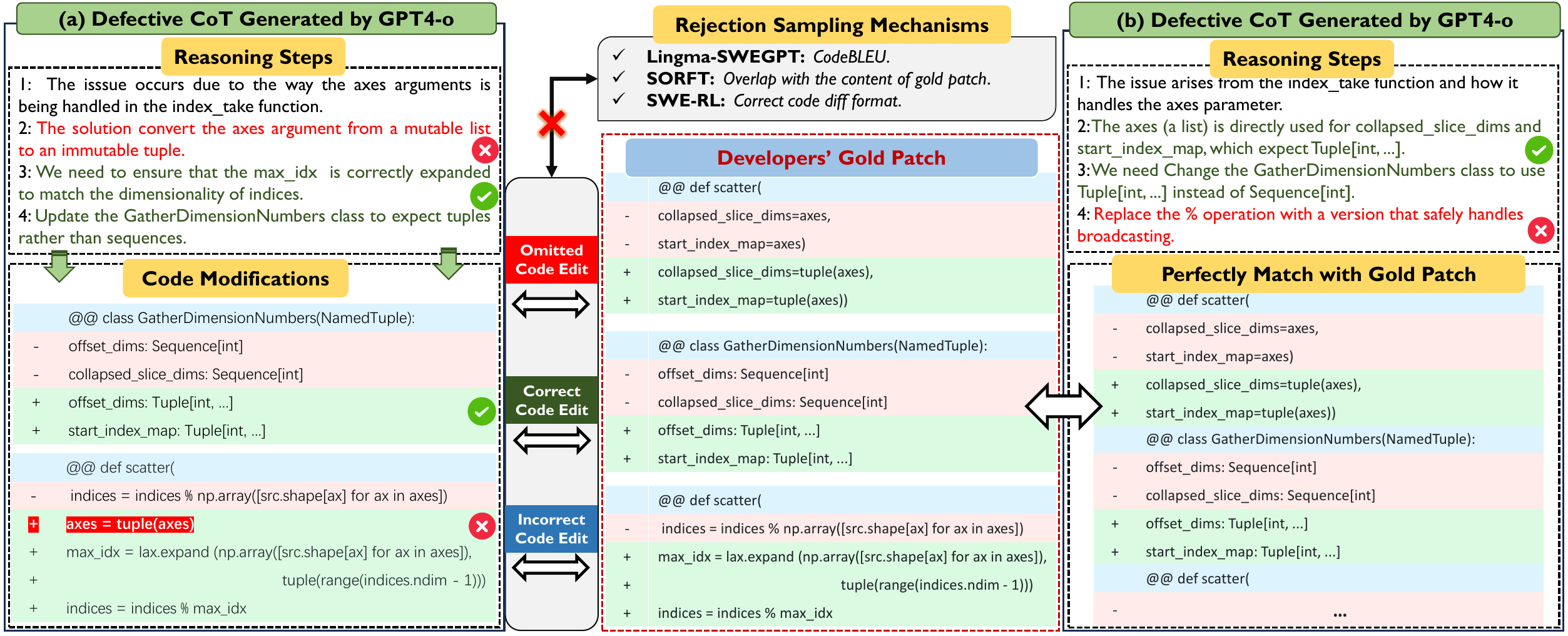}
	\vspace{-1mm}
	\caption{A illustrative example of defective CoT generated by \emph{GPT-4o}}
	\label{exa}
	\vspace{-6mm}
\end{figure*}

\subsection{\textbf{Limitation 1:} Inadequate Rejection Sampling Mechanisms for CoT Generation}
\label{Limitation 1}

To ensure the high-quality of CoT data generation,  prior studies~\cite{SWE-GPT, SWE-Gym, Sorft, SWE-RL} implements \textbf{rejection sampling with validation metrics} to filter out CoT instances that deviate from developer-approved issue patch records (\emph{\textbf{ground truth}}):

\begin{itemize}[leftmargin=*, topsep=1pt, itemsep=1pt]
	
\item \textit{\textbf{Lingma-SWE GPT }}~\cite{SWE-GPT}: Utilizing \emph{Jaccard Similarity} and \emph{CodeBLEU metrics} to eliminate low-quality CoT data that exhibits significant semantic deviation from ground truth in both issue localization and patch generation sub-tasks.

\item \textit{\textbf{\textsc{Sorft} }}~\cite{Sorft}: Filtering out CoT data that exhibit no overlap with ground truth specifications in either: (1) the actual patch locations (file names and line numbers), or (2) the code content of gold patches.

\item \textit{\textbf{SWE-RL }}~\cite{SWE-RL}: Filtering out CoT data that violates code diff formatting requirements, without verifying the actual correctness of the reasoning chains.

\item \textit{\textbf{SWE-Gym}}~\cite{SWE-Gym}: Validating CoT correctness through unit test execution and eliminating failing instances. While this improves CoT data quality, the per-instance test environment requirement makes it prohibitively expensive to scale.

\item \textit{\textbf{SWE-Fixer }}~\cite{SWE-Fixer}: The CoT data received no validation of its reasoning steps or final outputs.

\end{itemize}

However, the rejection sampling mechanisms proposed by the existing issue resolution approaches~\cite{SWE-GPT, SWE-Gym, Sorft, SWE-RL} cannot ensure the correctness of filtered CoT data. As illustrated in Figure~\ref{exa}(a), even when prompting \emph{GPT-4o} with \emph{issue descriptions}, \emph{target code}, and \emph{developers' gold patches}, the generated CoT data still contain errors.
Although the CoT contains the correct modification steps (e.g., changing {\mycode offset\_dims} to a {\mycode tuple} type), it also includes incorrect operations not present in the gold patch (such as converting the type of {\mycode axes} to a {\mycode tuple}).
Additionally, the CoT omits a critical modification from the ground truth (assigning axes to {\mycode start\_index\_map} after converting it to a {\mycode tuple} type). Existing rejection sampling mechanisms that use validation metrics like \emph{Jaccard Similarity}, \emph{CodeBLEU}, \emph{overlap with ground truth patches}, or \emph{code diff formatting requirements} fail to effectively filter out such defective CoT data.

\subsection{\textbf{Limitation 2:} Lack of Systematic Validation for Intermediate Reasoning Steps in CoT}
\label{Limitation 2}

Existing works~\cite{SWE-GPT, SWE-Gym, Sorft, SWE-RL} typically adopt a ``single-turn Q\&A'' approach, where LLMs are prompted with information such as \emph{issue descriptions} and \emph{repository context} to generate CoT for issue resolution. However, these works only validate the final outputs while failing to detect and correct errors in the intermediate reasoning steps of the CoT. This approach may cause the LLMs to exhibit ``error propagation'' during CoT generation, where mistakes in earlier steps lead to subsequent reasoning deviating from the correct trajectory. Particularly in repository-level issue resolution, such errors tend to accumulate and amplify, ultimately compromising the quality of the CoT data.

As illustrated in Figure~\ref{exa}(b), even when the LLM's output perfectly matches the developer's gold patch, errors may persist in the CoT's intermediate reasoning steps. 
The absence of dynamic validation and correction mechanisms for intermediate reasoning steps can significantly reduce the success rate of generating accurate CoT data.
More critically, enforcing a strict ``\emph{exact match with gold patch}'' validation metric to filter ``single-turn Q\&A''-generated CoT data—without intermediate reasoning corrections—would severely limit valid data retention, undermining fine-tuning effectiveness.

\section{\textsc{MCTS-Refine} Approach}

\emph{\textbf{Insight:}} ``Single-turn Q\&A prompting''—providing LLMs with issue descriptions and repository context to generate issue resolution CoT data—lacks validation mechanisms for intermediate reasoning steps. The absence of rigorous rejection sampling compromises intermediate reasoning step quality, while requiring \emph{exact ground truth alignment} severely limits retained high-quality CoT instances under ``single-turn Q\&A prompting'' approach, ultimately undermining model training effectiveness.
Monte Carlo Tree Search (MCTS)~\cite{MCTS} presents a viable alternative by efficiently navigating complex solution spaces. Unlike ``single-turn QA-based CoT generation'', MCTS enables (1) hierarchical decision tree exploration and (2) dynamic selection of optimal reasoning paths through iterative simulation. Yet traditional MCTS faces two key challenges: (1) error propagation from unverified intermediate steps, and (2) a reward-maximization strategy may fail to identify correct reasoning paths when all candidate paths contain flaws.

\begin{figure*}[]
	\centering
    \vspace{-3mm}
\includegraphics[width=0.85\textwidth]{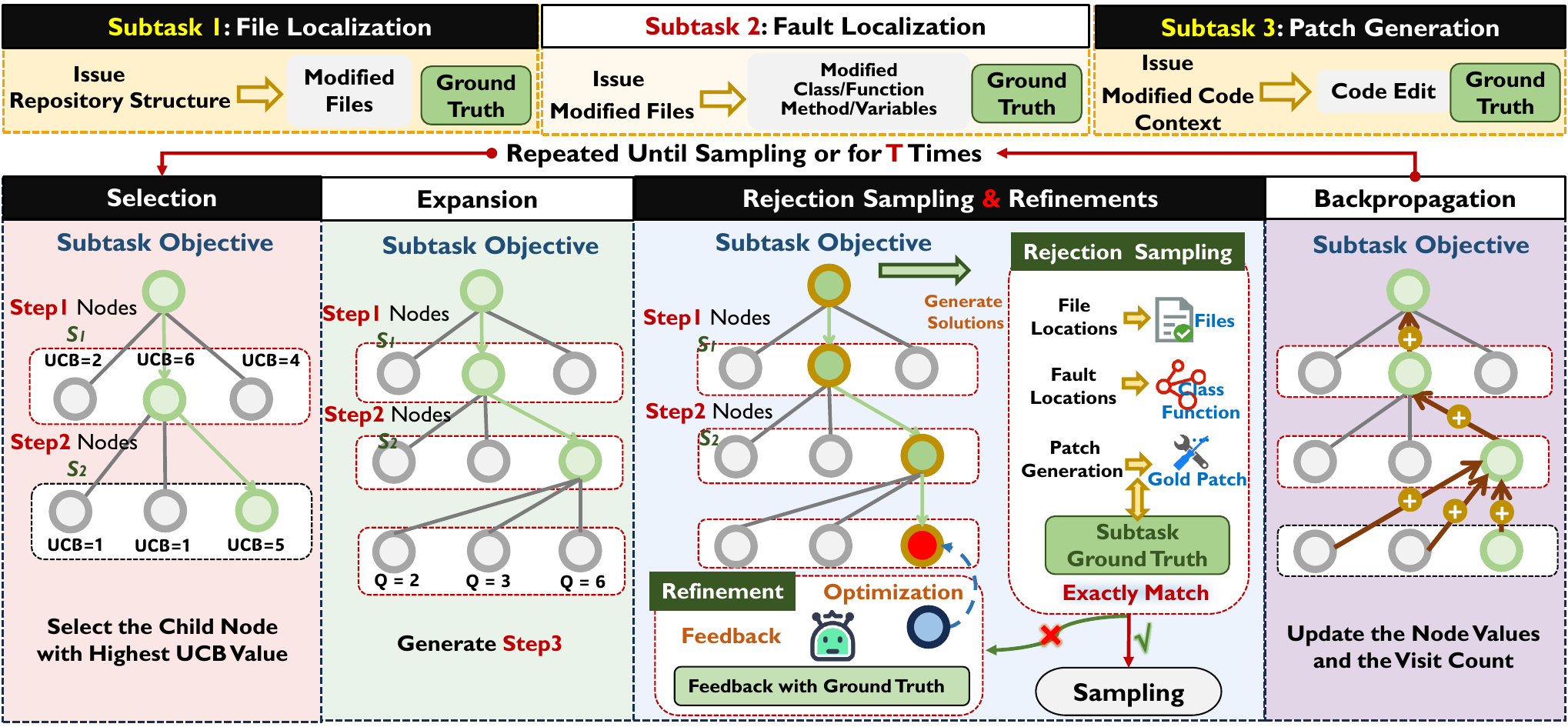}
	\vspace{-2mm}
	\caption{An Overall Architecture of \textsc{MCTS-REFINE}}
	\label{overview}
	\vspace{-6mm}
\end{figure*}

\emph{\textbf{Architecture:}} To address the aforementioned issues, we propose \textsc{MCTS-Refine}, an enhanced approach that integrates \textbf{CoT sampling} and \textbf{intermediate step reflection mechanism} into MCTS. Figure~\ref{overview} illustrates the architecture of \textsc{MCTS-Refine}, which consists of five core components:
(1) \emph{\textbf{Selection}}: Chooses the current optimal node using the Upper Confidence Bound (UCB) strategy.
(2) \emph{\textbf{Expansion}}: Generates new reasoning steps and assigns reward values.
(3) \emph{\textbf{Rejection Sampling}}: Validates whether the reasoning path can correctly solve the task.
(4) \emph{\textbf{Refinement}}: Provides feedback and corrects erroneous steps in the reasoning path.
(5) \emph{\textbf{Backpropagation}}: Updates node reward values and visit counts to optimize the entire reasoning path.
It is worth emphasizing that:

\begin{itemize}[leftmargin=*, topsep=1pt, itemsep=1pt]
	
\item \textit{\textbf{Rejection Sampling}} component employs a strict ``exact match'' mechanism against ground truth to verify the validity of the current reasoning path.

\item \textit{\textbf{Refinement}} component performs two tasks: (1) detecting deviations from ground truth in current reasoning steps while providing corrective suggestions, and (2) optimizing erroneous steps by incorporating these modifications.

\end{itemize}

To address the complex issue resolution tasks, we adopt the \textsc{Agentless} framework~\cite{Agentless} by decomposing the process into three subtasks: \emph{\textbf{File Localization}}, \emph{\textbf{Fault Localization}}, and \emph{\textbf{Patch Generation}}. This structured decomposition enables: (1) \emph{\textbf{clear definition of input-output specifications for each subtask}}; (2) \emph{\textbf{systematic construction of ground truth datasets for each subtask phase}}; (3) \emph{\textbf{generate high-quality CoT data specifically tailored for each subtask leveraging \textsc{MCTS-Refine}}}.
This approach enhances the model's reasoning performance across all subtasks in the issue resolution pipeline.

\subsection{Task Decomposition for Issue Resolution}
Task decomposition in issue resolution effectively reduces complexity while clarifying objectives for individual components. Our approach partitions the issue resolution process into three distinct subtasks, employing \textsc{MCTS-Refine} to generate high-quality CoT reasoning for each. In contrast to conventional relaxed sampling mechanism adopted by existing approaches~\cite{SWE-GPT, SWE-Gym, SWE-Fixer, Sorft, SWE-RL}, \textsc{MCTS-Refine} implements a rigorous stepwise generation process with a ground truth-aligned rejection sampling criterion that guarantees reasoning path validity.

\emph{\textbf{File Localization:}} This phase focuses on identifying target files for modification based on \emph{issue descriptions} and \emph{repository structure}. During CoT data sampling, we process both the issue description and repository structure as inputs. \textsc{MCTS-Refine} progressively generates and explores reasoning paths, accepting CoT sequences only when their resulting file localization matches the actual files modified by developers.

\emph{\textbf{Fault Localization:}} This phase focuses on identifying issue-introducing code elements (e.g., classes, methods, functions, and global variables), building upon \emph{\textbf{File Localization}}. The CoT generation process takes the issue description and previously localized file structure as inputs, retaining only those reasoning paths whose predicted fault locations exactly match the actual code positions modified in developers' gold patches.

\emph{\textbf{Patch Generation:}} This phase empowers the LLM to generate precise code fixes for the identified issues by processing both the issue description and fault-localized code elements along with their relevant contextual code segments from the previous \emph{\textbf{Fault Localization}} phase. During CoT construction, the system validates a reasoning path only when the LLM-generated patch achieves exact match with developers' gold patch. To eliminate confounding factors, we: (1) integrate the LLM's code edits back into the original repository context; (2) normalize the output by removing comments, whitespace, and line breaks; (3) perform strict differential comparison against the gold patch. This rigorous verification protocol ensures the training data's authenticity by enforcing syntactic and functional equivalence with developer solutions.

\subsection{MCTS With Rejection Sampling and Refinement}
Current ``single-turn QA-based CoT generation'' approaches lack systematic validation of intermediate reasoning steps, preventing dynamic path adjustment and often yielding suboptimal solutions with defective CoT data. While traditional MCTS algorithms can dynamically narrow the search space by simulating multiple reasoning trajectories (through \emph{Selection}, \emph{Expansion}, \emph{Simulation}, and \emph{Backpropagation} phases), they exhibit two critical flaws: (1) During simulation, MCTS only estimates node potential without verifying step correctness, allowing error propagation in multi-step reasoning. (2) The threshold-based path selection cannot guarantee the final solution’s validity.
\textsc{MCTS-Refine} addresses these limitations by: (1) Introducing a CoT sampling strategy for real-time correct path identification; (2) Incorporating a reflection mechanism to rectify erroneous reasoning steps.

\textsc{MCTS-Refine} fundamentally models the reasoning process as a search tree structure, where: \emph{\textbf{the root node encapsulates the subtask objective, and each subsequent node represents an individual reasoning step.}} The algorithm iteratively refines solutions through five core phases: \emph{\textbf{Selection}}, \emph{\textbf{Expansion}}, \emph{\textbf{Rejection Sampling}}, \emph{\textbf{Refinement}}, and \emph{\textbf{Backpropagation}}. Below we elaborate on the phase-specific implementations.

\noindent\emph{{\textbf{B.1. Selection}}}
{~}

During the \emph{\textbf{selection phase}}, \textsc{MCTS-Refine} balances exploration and exploitation by traversing the tree using a predefined policy until it reaches a not fully explored node. This policy weighs both the node's reward value and visit count, ensuring a preference for high-reward nodes while still allowing exploration of less-visited ones—thus preventing convergence to local optima.

In the issue resolution task, the \textbf{root node} corresponds to the subtask objective (e.g., precisely identifying target files in \emph{File Localization}), while \textbf{child nodes} encode all potential reasoning steps. The \textbf{leaf nodes} constitute the search boundary, comprising either: unexplored nodes awaiting expansion, or terminal nodes representing verified solutions. Each iteration initiates at the root, then recursively applies \emph{UCB}-based selection to navigate through child nodes until encountering a leaf node. This traversal path's node sequence then provides contextual grounding for the subsequent \emph{\textbf{expansion phase}}.



\begin{equation}
\footnotesize
UCB(S) = Q_S + \epsilon \cdot \sqrt{\frac{\ln N_{\text{parent}}}{N_S}}
\label{eq:ucb}
\end{equation}

In this formulation, the reward score $Q_S$ for node $S$ is assigned by the LLM during node expansion, where $N_{parent}$ and $N_S$ represent the visit counts of the parent node and current node $S$ respectively. The exploration parameter $\epsilon$ is set to 0.5 following established practice~\cite{Rest-mcts}, balancing exploration of new nodes against exploitation of known high-reward paths.


\noindent\emph{{B.2. \textbf{Expansion}}}
{~}

\begin{figure}[]
	\centering
	\vspace{-2mm}
\includegraphics[width=0.43\textwidth]{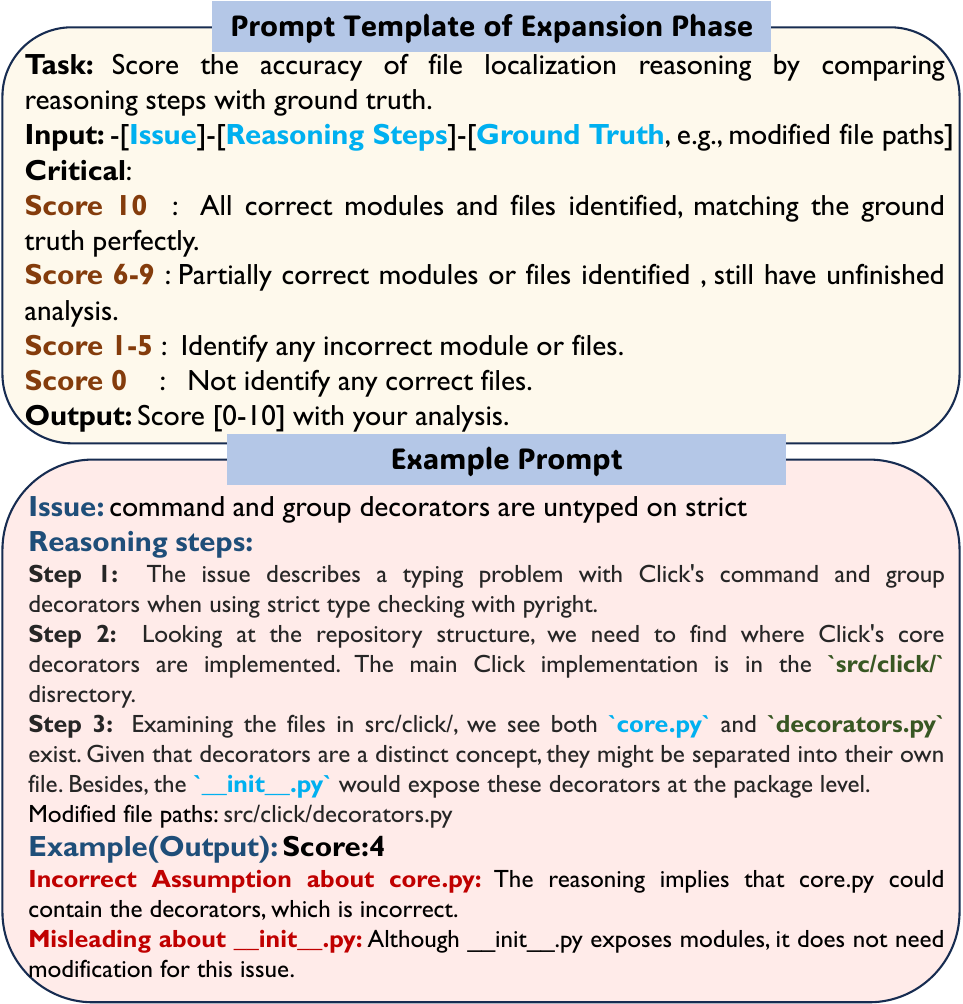}
	\caption{An illustrative Prompt Template of \emph{\textbf{Expansion Phase}} for \emph{\textbf{File Localization}} Subtask}
	\label{prompt1}
	\vspace{-5mm}
\end{figure}

The \textbf{\emph{expansion phase}} creates new child nodes from the leaf node generated in the \textbf{\emph{selection phase}}, with each child node corresponding to a unique reasoning step. Formally, let $S_k$ represent the current selected node (denoting the $k$-\emph{th} reasoning step), and $P_k$ = ($S_1$, $S_2$, $\cdots$, $S_k$) define the complete reasoning path from the root node. \textsc{MCTS-Refine} extends path $P_k$ by creating $b$ new child nodes $S_{k+1}$, where we adopt the conventional branching factor of $b$=3 as established in prior MCTS literature~\cite{Rest-mcts}.

The LLM assigns a reward score to each newly generated child node to evaluate the validity of its corresponding reasoning step. This assessment follows a rigorous process: (1) inputting the current reasoning path and subtask ground truth to the LLM, (2) analyzing the deviation between them using predefined scoring metrics in the prompt template, and (3) generating a reward score on a \textbf{0-10 scale}, where higher values indicate closer alignment with the correct solution. We have developed specialized prompt templates for each subtask to ensure consistent quality assessment of reasoning paths.

As depicted in Figure~\ref{prompt1}, during \textbf{\emph{File Localization}}, the LLM assigned a reward score of 4 to the current reasoning path based on the provided inputs. This suboptimal score resulted from a critical error in \emph{Step 3}, where the LLM incorrectly identified both {\mycode core.py} and {\mycode \_init\_.py} files, leading to substantial deviation from the expected reasoning path.

\noindent\emph{{B.3. \textbf{Rejection Sampling}}}
{~}

In the classical MCTS algorithm, after \textbf{\emph{expansion phase}}, a node is selected for simulation based on its \emph{UCB} value. While this simulation phase aims to estimate the node's potential value, it cannot verify whether the current reasoning path can fully solve the subtask—particularly when the path contains flawed reasoning steps. This limitation arises because each node only stores partial reasoning steps toward task completion without generating the actual subtask solution, potentially leading to unnecessary exploration of incorrect nodes in later iterations.
To overcome this issue, we introduce CoT sampling and reflection immediately after node expansion. This enhancement ensures that: (1) proactively identifies valid reasoning paths, and (2) detects and corrects flawed reasoning steps during exploration.

The mechanism evaluates reasoning paths from the root node to its expanded optimal nodes by using an LLM to generate corresponding subtask solutions. These generated solutions are then compared against the ground truth to determine whether the current reasoning path should be sampled. The specific sampling rules are as follows:

\begin{itemize}[leftmargin=*, topsep=1pt, itemsep=1pt]
	
\item \textit{\textbf{File Location Subtask:}} Samples reasoning paths where the identified files exactly match the developer-modified files.

\item \textit{\textbf{Fault Localization Subtask:}} Samples reasoning paths where the identified class signatures, method signatures, function signatures, and global variables completely align with the developer modifications.

\item \textit{\textbf{Patch Generation Subtask:}} Integrates the generated edit code (based on the reasoning path) into the original repository and compares the modified files with the developer-edited versions. Our tool samples reasoning paths where the generated edits exactly match the ground truth. It should be noted that the comparison ignores whitespace, line breaks, and comments. 
We intentionally avoid direct comparisons between the generated edits and the gold patch. This is because the gold patch only contains modified sections with limited context, whereas the LLM's output typically includes both modifications and additional contextual code. This approach effectively prevents potential false negatives in our evaluation arising from the inherent structural disparities between these two distinct formats.
	
\end{itemize}

\noindent\emph{{B.4. \textbf{Refinement}}}
{~}

The \emph{\textbf{refinement phase}} serves to identify and rectify potential errors in current reasoning steps through two coordinated sub-phases: the \emph{\textbf{Feedback sub-phase}} detects reasoning errors and generates diagnostic feedback, while the \emph{\textbf{Optimization sub-phase}} subsequently performs corrective modifications and optimizations based on this feedback.

\begin{equation}
F_{k+1} = Feedback(P_k, GT)
\label{eq:F}
\end{equation}

\begin{figure}[]
	\centering
	\vspace{-8mm}
\includegraphics[width=0.43\textwidth]{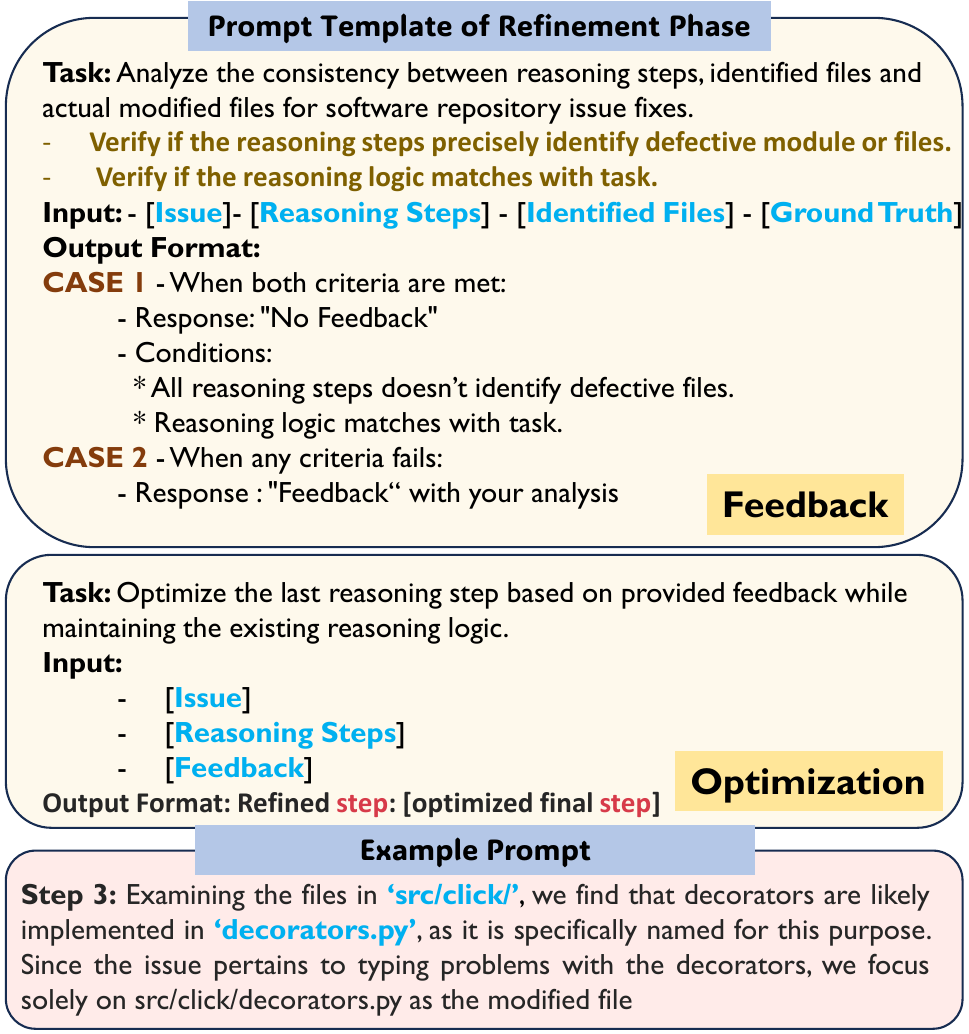}
	\caption{An illustrative Prompt Template of \emph{\textbf{Refinement Phase}} for \emph{\textbf{File Localization}} Subtask}
	\label{prompt2}
	\vspace{-6mm}
\end{figure}

During the \emph{\textbf{Feedback sub-phase}}, the model generates feedback $F_{k+1}$ by evaluating the current reasoning path $P_k$ = \{$S_1$, $S_2$, $\cdots$, $S_{k+1}$\} against the ground truth ($GT$), as formalized in Equation (2). This phase executes two critical functions: (1) assessing the alignment between the reasoning path and ground truth, and (2) generating specific corrective suggestions for identified deviations.
Figure~\ref{prompt2} shows the prompt template of \emph{\textbf{Feedback sub-phase}}, which produces two distinct outputs:

\begin{itemize}[leftmargin=*, topsep=1pt, itemsep=1pt]
	
\item \emph{\textbf{No-Feedback:}} Indicates full path correctness requiring no modification.

\item \emph{\textbf{Feedback:}} Provides specific corrective instructions when errors are detected. For instance, in \emph{\textbf{File Localization}} subtasks, the feedback mechanism evaluates whether the current path includes directories or files erroneously identified by the developer (as specified in the ground truth). As demonstrated in Figure~\ref{prompt2}'s prompt template, when the path incorrectly locates both {\mycode core.py} and \texttt{\mycode \_init\_.py} files, the model not only flags these inaccuracies but also generates concrete revision guidance. Similarly, for \emph{\textbf{Patch Generation}} subtasks, the feedback verifies both the syntactic correctness of modified code segments and their semantic alignment with the ground truth's modification strategy.

\end{itemize}

Upon receiving corrective \emph{\textbf{Feedback}}, the system initiates the \emph{\textbf{Optimization sub-phase}} to refine the reasoning path. The LLM dynamically adjusts the current reasoning step $S_{k+1}$ using the diagnostic feedback and optimizes the reasoning path. This process can be formally represented as:

\begin{equation}
\tilde{S}_{k+1} = Optimization (S_{k+1}, F_{k+1})
\label{eq:Op}
\end{equation}

\noindent where $\tilde{S}_{k+1}$ represents the refined reasoning step after optimization. This phase focuses exclusively on improving the current step while deliberately restricting next-step generation. As illustrated in Figure~\ref{prompt2}, the LLM successfully rectifies flawed reasoning steps by integrating reflective analysis with current step evaluation. This constrained correction approach ensures: (1) \emph{\textbf{Localized Modification Scope:}} Each adjustment is precisely bounded to the target step. (2) \emph{\textbf{Global Path Stability:}} Preserves the structural integrity of the overall reasoning chain. (3) \emph{\textbf{Controlled Optimization:}} Prevents cascading disruptions across the reasoning path.

\noindent\emph{{B.5. \textbf{Backpropagation}}}
{~}

During the \emph{\textbf{Backpropagation phase}}, we perform a bottom-up update from the selected node $S_{k+1}$ back to the root node. This sequential upward propagation ensures that all nodes along the path are adjusted according to the latest reward values.
The key objective of this phase is to update: (1) the visit count $N_S$, and (2) 
the reward value $Q_S$, for each reasoning step node $S_i$ in the path $P_k$ = \{$S_1$, $S_2$, $\cdots$, $S_{k+1}$\}.
The update process involves: (1) incrementing the visit count $N_{S_i}$ of each node $S_i$ by 1. (2) updating the reward value $Q_{S_i}$ based on the reward contributions from $b$ child nodes $S_{i+1}$ of $S_i$.

\begin{equation}
Q_{increment} = \frac{\sum_{i=1}^{b} N_{S_{i+1}} \times Q_{S_{i+1}}}{\sum_{i=1}^{b}  N_{S_{i+1}}}
\end{equation}

\begin{equation}
Q_{S_i} = \alpha \times Q_{S_i} + (1 - \alpha) \times Q_{increment}
\end{equation}

The weighting factor $\alpha$ is set to 0.5 to balance between: (1) the parent node's current reward value $Q_{S_i}$, and (2) the reward increments $Q_{increment}$ from its 
$b$ child nodes $S_{i+1}$. Through this update rule, $Q_{S_i}$ progressively incorporates child-node rewards, resulting in a more accurate representation of the node’s importance in the reasoning path $P_k$.

Repeat the process (\emph{\textbf{selection}} $\rightarrow$ \emph{\textbf{expansion}} $\rightarrow$ \emph{\textbf{rejection sampling}} $\rightarrow$ \emph{\textbf{refinement}} $\rightarrow$ \emph{\textbf{backpropagation}}) until a reasoning path and result consistent with the ground truth are sampled, or terminate after \textit{T} iterations.





\subsection{Implementation}

\emph{\textbf{Dataset Construction:}} To construct \textsc{MCoT} dataset for the issue resolution task, we leverage the \emph{SWE-Fixer-Train-110K} dataset collected by \textsc{SWE-Fixer}~\cite{SWE-Fixer}, comprising 110,000 high-quality Issue and Pull Request (PR) pairs sourced from GitHub repositories (excluding those overlapping with \textsc{SWE-Bench}). For effective subtask decomposition, we utilized the GitHub REST API to enrich each Issue-PR pair with contextual information including: (1) repository directory structures, and (2) pre-patch code file contents, thereby creating comprehensive inputs for each subtask's processing pipeline.


\emph{\textbf{Inference Framework:}} We leverage our constructed CoT data \textsc{MCoT} to fine-tune LLMs across multiple parameter scales and integrate the resulting models into the \textsc{Agentless 1.0}~\cite{Agentless} framework. This framework operates through two distinct phases: (1) \emph{Fault Localization}: The LLM hierarchically narrows down potential issue-introducing candidates (files $\rightarrow$ classes $\rightarrow$ functions) to pinpoint issue-relevant code segments. (2) \emph{Code Edit Generation}: The LLM then produces syntactically valid code modifications to resolve the identified issues. The \textsc{Agentless 1.0} framework's architecture is generally compatible with our CoT data, enabling end-to-end evaluation of \textsc{MCTS-Refine}'s \emph{Rejection Sampling} and \emph{Refinement} methodology.
\section{Evaluation}

In evaluation section, we propose three research questions:

\textbf{RQ1: (\emph{Overall Effectiveness of Issue Resolution}):} \emph{To what extent does the fine-tuned model improve overall performance in issue resolution tasks compared to SOTA baselines?}

\textbf{RQ2: (\emph{Effectiveness of File Location and Fault Localization}):} \emph{What is the performance of the fine-tuned model on the File Localization and Fault Localization subtasks?} 

\textbf{RQ3: (\emph{Effectiveness of Patch Generation}):} \emph{What is the performance of the fine-tuned model on the Patch Generation subtask?}

\subsection{Experiment Setup}

\subsubsection{\textbf{CoT Data Synthesis}} To reduce computational overhead and experimental costs, we randomly selected 20,000 Issue-PR pairs from \emph{SWE-Fixer-Train-110K}\cite{SWE-Fixer} to build our CoT dataset using \textsc{MCTS-Refine}, for three core issue resolution subtasks—\emph{\textbf{File Localization}}, \emph{\textbf{Fault Localization}}, and \emph{\textbf{Patch Generation}}. 
The data synthesis employed \emph{DeepSeek-v3}, which provides three advantages over \emph{GPT-4o} and \emph{Claude-3.5-Sonnet}: (1) \emph{lower computational costs}, (2) \emph{complete open-source availability}, and (3) \emph{full local deployability}.

The refinement process iterates through six MCTS phases—\emph{\textbf{selection}}, \emph{\textbf{expansion}}, \emph{\textbf{rejection}}, \emph{\textbf{sampling}}, \emph{\textbf{refinement}}, and \emph{\textbf{backpropagation}}—terminating when either: (1) the derived solution perfectly matches the ground truth, or (2) reaches our empirically set maximum of 50 iterations ($T$ = 50) based on the existing MCTS approach\cite{Rest-mcts}. The final curated fine-tuning dataset comprises 52,068 high-quality samples with the following distribution: (1) \emph{\textbf{File Localization}}: 18,340 samples; (2) \emph{\textbf{Fault Localization}}: 17,338 samples; (3) \emph{\textbf{Patch Generation}}: 16,390 samples.

\subsubsection{\textbf{Training Configuration}} We fine-tuned three open-source LLMs of varying parameter scales using our constructed CoT data: \textit{Qwen2.5-Coder-7B-Instruct}, \textit{Qwen2.5-Coder-32B-Instruct}, and \textit{Qwen2.5-72B-Instruct}~\cite{qwenmodels}. These models achieve SOTA performance at their respective parameter scales and demonstrate robust code processing capabilities. For the 7B and 32B models, we conducted full-parameter supervised fine-tuning (SFT) using \emph{Llama-Factory}~\cite{zheng2024llamafactory}, completing training on eight {\mycode NVIDIA H800 80GB GPUs}. The training configuration included: (1) \emph{Context window length}: 32k tokens; (2) \emph{Batch size}: 128; (3) \emph{Training duration}: 2 epochs; (4) \emph{Learning rate}: 5e-6 initial rate with cosine decay; (5) \emph{Warmup ratio}: 3\%. Due to computational constraints, we adapted the 72B model using LoRA\cite{hu2022lora} fine-tuning via the \emph{Llama-Factory} framework while maintaining consistent hyperparameters.

\subsubsection{\textbf{Baselines}} Our evaluation systematically compares the fine-tuned model against both proprietary and open-source LLMs across different issue resolution frameworks.

\begin{itemize}[leftmargin=*, topsep=1pt, itemsep=1pt]

\item \emph{\textbf{Framework Selection:}} Our comparison considers two types of issue resolution frameworks: (1) \emph{\textbf{Agent-based frameworks}} (e.g., \textsc{Lingma-SWE-GPT}~\cite{SWE-GPT}, \textsc{Openhands}~\cite{Openhands}, \textsc{SWE-Agent}~\cite{Swe-agent}) that dynamically integrate external tools like code analyzers and compilers to enable adaptive, context-aware repair processes, and (2) \emph{\textbf{Pipeline-based frameworks}} (e.g., \textsc{SWE-Fixer}~\cite{SWE-Fixer}'s RAG-to-patch generation workflow, \textsc{Agentless}~\cite{Agentless}'s decomposition methodology) that employ structured, modular task sequencing with well-defined interfaces - with the former excelling in flexibility through on-demand tool invocation while the latter provides reproducible evaluation frameworks particularly valuable for benchmarking LLM capabilities.

\item \emph{\textbf{Model Selection:}} We comprehensively evaluated our fine-tuned LLMs against \emph{\textbf{cutting-edge proprietary models}} (including {\mycode OpenAI}'s \emph{GPT-4} and \emph{GPT-4o}~\cite{GPT-4o} - renowned for their exceptional natural language understanding and efficient inference capabilities, and {\mycode Anthropic}'s \emph{Claude 3 Opus} and \emph{Claude 3.5 Sonnet}~\cite{Claude-3.5-Sonnet} - particularly strong in coding tasks and multi-turn interactions) as well as \emph{\textbf{leading open-source models}} (\emph{Qwen} series models like \emph{Qwen2.5-Coder-7B-Instruct}~\cite{qwenmodels} that we fine-tuned, and \emph{Deepseek} models including the general-purpose \emph{Deepseek-V3}~\cite{liu2024Deepseek} and reasoning-specialized \emph{Deepseek-R1}~\cite{guo2025deepseekr1}), with all comparison models carefully selected from prominent research works and the \emph{SWE-bench} leaderboard to ensure a rigorous benchmark of our approach's effectiveness across different model categories and capabilities.

\end{itemize}

\subsubsection{\textbf{Benchmark and Metrics}} We conducted rigorous performance assessments using the industry-standard \textbf{\emph{SWE-bench}} benchmark suite, specifically leveraging both its \textbf{\emph{Verified}} (500 samples) and \textbf{\emph{Lite}} (300 samples) subsets comprising real-world GitHub issues~\cite{SWE-Bench}. \emph{SWE-bench} has become the authoritative benchmark for issue resolution tasks, now serving as the standard evaluation framework for leading AI coding systems.

We analyzed model performance following the \textsc{Agentless}~\cite{Agentless} framework's workflow, focusing on four metrics: \textbf{(1) \emph{\%Resolved:}} \emph{Success rate of model-generated patches passing all tests.} \textbf{(2) \emph{\%FileHit:}} \emph{Precision in file-level localization.} \textbf{(3) \emph{\%FuncHit:}} \emph{Precision in class/method-level localization.} \textbf{(4) \emph{\%Line-Hit:}} \emph{Precision in code line-level localization.}

\subsection{RQ1: Overall Effectiveness of Issue Resolution}
\noindent\textbf{\emph{Methodology.}} To comprehensively evaluate the performance of the fine-tuned model in resolving real-world GitHub issues, we conducted comparative experiments with baselines based on \emph{SWE-bench Verified} and \emph{SWE-bench Lite}. Our evaluation protocol strictly provided only: (1) the original issue description and (2) corresponding repository context as model inputs, with patch correctness being definitively determined through test-case verification. 
 We measured the model's performance by quantifying the number of issues resolved (\emph{\textbf{\%Resolved}}).

\begin{table}[]
\scriptsize
\centering
\setlength\tabcolsep{1pt}
\def\arraystretch{1}
\vspace{-5mm}
\caption{Overall Effectiveness of Issue Resolution}
\vspace{-2mm}
 \bgroup
\begin{tabular}{|cccccc|}
\hline
\rowcolor[HTML]{343434} 
\multicolumn{1}{|c|}{\cellcolor[HTML]{343434}{\color[HTML]{FFFFFF} \textit{Framework}}} & \multicolumn{1}{c|}{\cellcolor[HTML]{343434}{\color[HTML]{FFFFFF} \textit{Models}}} & \multicolumn{1}{c|}{\cellcolor[HTML]{343434}{\color[HTML]{FFFFFF} \textit{SFT}}} & \multicolumn{1}{c|}{\cellcolor[HTML]{343434}{\color[HTML]{FFFFFF} \textit{Type}}} & \multicolumn{1}{c|}{\cellcolor[HTML]{343434}{\color[HTML]{FFFFFF} \textit{Verified}}} & {\color[HTML]{FFFFFF} \textit{ Lite}} \\ \hline \hline
\rowcolor[HTML]{EFEFEF} 
\multicolumn{6}{|c|}{\cellcolor[HTML]{EFEFEF}\textit{\textbf{Proprietary Models}}}                                                                                                                                                                                                                                                                                                                                                                                                               \\ \hline
\rowcolor[HTML]{FFFFFF} 
\multicolumn{1}{|c|}{\cellcolor[HTML]{FFFFFF}\textit{OpenHands}~\cite{Openhands}}                        & \multicolumn{1}{c|}{\cellcolor[HTML]{FFFFFF}\textit{GPT4-o}}                        & \multicolumn{1}{c|}{\cellcolor[HTML]{FFFFFF}\emph{\text{N/A}}}                                   & \multicolumn{1}{c|}{\cellcolor[HTML]{FFFFFF}\textit{Agent}}                       & \multicolumn{1}{c|}{\cellcolor[HTML]{FFFFFF}-}                                                    & 22.0\%                                    \\ \hline
\rowcolor[HTML]{FFFFFF} 
\multicolumn{1}{|c|}{\cellcolor[HTML]{FFFFFF}\textit{OpenHands}~\cite{Openhands}}                        & \multicolumn{1}{c|}{\cellcolor[HTML]{FFFFFF}\textit{Claude-3.5-Sonnet}}    & \multicolumn{1}{c|}{\cellcolor[HTML]{FFFFFF}\emph{\text{N/A}}}                                   & \multicolumn{1}{c|}{\cellcolor[HTML]{FFFFFF}\textit{Agent}}                       & \multicolumn{1}{c|}{\cellcolor[HTML]{FFFFFF}53.0\%}                                                 & 41.7\%                                    \\ \hline
\rowcolor[HTML]{FFFFFF} 
\multicolumn{1}{|c|}{\cellcolor[HTML]{FFFFFF}\textit{SWE-Agent}~\cite{Swe-agent}}                        & \multicolumn{1}{c|}{\cellcolor[HTML]{FFFFFF}\textit{Claude-3-Opus}}                 & \multicolumn{1}{c|}{\cellcolor[HTML]{FFFFFF}\emph{\text{N/A}}}                                   & \multicolumn{1}{c|}{\cellcolor[HTML]{FFFFFF}\textit{Agent}}                       & \multicolumn{1}{c|}{\cellcolor[HTML]{FFFFFF}18.2\%}                                                 & 11.7\%                                    \\ \hline
\rowcolor[HTML]{FFFFFF} 
\multicolumn{1}{|c|}{\cellcolor[HTML]{FFFFFF}\textit{SWE-Agent}~\cite{Swe-agent}}                        & \multicolumn{1}{c|}{\cellcolor[HTML]{FFFFFF}\textit{GPT-4o}}                        & \multicolumn{1}{c|}{\cellcolor[HTML]{FFFFFF}\emph{\text{N/A}}}                                   & \multicolumn{1}{c|}{\cellcolor[HTML]{FFFFFF}\textit{Agent}}                       & \multicolumn{1}{c|}{\cellcolor[HTML]{FFFFFF}23.0\%}                                                 & 18.3\%                                    \\ \hline
\rowcolor[HTML]{FFFFFF} 
\multicolumn{1}{|c|}{\cellcolor[HTML]{FFFFFF}\textit{SWE-Agent}~\cite{Swe-agent}}                        & \multicolumn{1}{c|}{\cellcolor[HTML]{FFFFFF}\textit{Claude-3.5-Sonnet}}             & \multicolumn{1}{c|}{\cellcolor[HTML]{FFFFFF}\emph{\text{N/A}}}                                   & \multicolumn{1}{c|}{\cellcolor[HTML]{FFFFFF}\textit{Agent}}                       & \multicolumn{1}{c|}{\cellcolor[HTML]{FFFFFF}33.6\%}                                                 & 23.0\%                                    \\ \hline
\rowcolor[HTML]{FFFFFF} 
\multicolumn{1}{|c|}{\cellcolor[HTML]{FFFFFF}\textit{Agentless}~\cite{Agentless}}                        & \multicolumn{1}{c|}{\cellcolor[HTML]{FFFFFF}\textit{GPT-4o}}                        & \multicolumn{1}{c|}{\cellcolor[HTML]{FFFFFF}\emph{\text{N/A}}}                                   & \multicolumn{1}{c|}{\cellcolor[HTML]{FFFFFF}\textit{Pipeline}}                    & \multicolumn{1}{c|}{\cellcolor[HTML]{FFFFFF}38.8\%}                                                 & 32.0\%                                    \\ \hline
\rowcolor[HTML]{FFFFFF} 
\multicolumn{1}{|c|}{\cellcolor[HTML]{FFFFFF}\textit{Agentless}~\cite{Agentless}}                        & \multicolumn{1}{c|}{\cellcolor[HTML]{FFFFFF}\textit{Claude-3.5-Sonnet}}    & \multicolumn{1}{c|}{\cellcolor[HTML]{FFFFFF}\emph{\text{N/A}}}                                   & \multicolumn{1}{c|}{\cellcolor[HTML]{FFFFFF}\textit{Pipeline}}                    & \multicolumn{1}{c|}{\cellcolor[HTML]{FFFFFF}50.8\%}                                                 & 40.7\%                                    \\ \hline
\rowcolor[HTML]{EFEFEF} 
\multicolumn{6}{|c|}{\cellcolor[HTML]{EFEFEF}\textit{\textbf{Open-source   Frameworks \& Open-source Models \textbf{(<10B)}}}}                                                                                                                                                                                                                                                                                                                                                                                \\ \hline
\rowcolor[HTML]{FFFFFF} 
\multicolumn{1}{|c|}{\cellcolor[HTML]{FFFFFF}\textit{Openhands}~\cite{Openhands}}                        & \multicolumn{1}{c|}{\cellcolor[HTML]{FFFFFF}\textit{SWE-Gym-Qwen-7B}}               & \multicolumn{1}{c|}{\cellcolor[HTML]{FFFFFF}\textit{SFT}}                        & \multicolumn{1}{c|}{\cellcolor[HTML]{FFFFFF}\textit{Agent}}                       & \multicolumn{1}{c|}{\cellcolor[HTML]{FFFFFF}10.6\%}                                                 & 10.0\%                                    \\ \hline
\rowcolor[HTML]{FFFFFF} 
\multicolumn{1}{|c|}{\cellcolor[HTML]{FFFFFF}\textit{Agentless}~\cite{Agentless}}                        & \multicolumn{1}{c|}{\cellcolor[HTML]{FFFFFF}\textit{SoRFT-Qwen-7B}}                 & \multicolumn{1}{c|}{\cellcolor[HTML]{FFFFFF}\textit{RL}}                         & \multicolumn{1}{c|}{\cellcolor[HTML]{FFFFFF}\textit{Pipeline}}                    & \multicolumn{1}{c|}{\cellcolor[HTML]{FFFFFF}\textcolor{blue}{21.4\%}}                                                 & \textcolor{blue}{14.0\%}                                    \\ \hline
\rowcolor[HTML]{FFFFFF} 
\multicolumn{1}{|c|}{\cellcolor[HTML]{FFFFFF}\textit{Lingma-SWE-GPT}~\cite{SWE-GPT}}                          & \multicolumn{1}{c|}{\cellcolor[HTML]{FFFFFF}\textit{Lingma-SWE-GPT-7B}}             & \multicolumn{1}{c|}{\cellcolor[HTML]{FFFFFF}\textit{SFT}}                        & \multicolumn{1}{c|}{\cellcolor[HTML]{FFFFFF}\textit{Agent}}                       & \multicolumn{1}{c|}{\cellcolor[HTML]{FFFFFF}18.4\%}                                                 & 12.0\%                                    \\ \hline
\rowcolor[HTML]{EFEFEF} 
\multicolumn{6}{|c|}{\cellcolor[HTML]{EFEFEF}\textit{\textbf{Open-source Frameworks \& Open-source Models \textbf{(>10B)}}}}                                                                                                                                                                                                                                                                                                                                                                                \\ \hline
\rowcolor[HTML]{FFFFFF} 
\multicolumn{1}{|c|}{\cellcolor[HTML]{FFFFFF}\textit{Openhands}~\cite{Openhands}}                        & \multicolumn{1}{c|}{\cellcolor[HTML]{FFFFFF}\textit{SWE-Gym-Qwen-14B}}              & \multicolumn{1}{c|}{\cellcolor[HTML]{FFFFFF}\textit{SFT}}                        & \multicolumn{1}{c|}{\cellcolor[HTML]{FFFFFF}\textit{Agent}}                       & \multicolumn{1}{c|}{\cellcolor[HTML]{FFFFFF}16.4\%}                                                 & 12.7\%                                    \\ \hline
\rowcolor[HTML]{FFFFFF} 
\multicolumn{1}{|c|}{\cellcolor[HTML]{FFFFFF}\textit{Agentless}~\cite{Agentless}}                        & \multicolumn{1}{c|}{\cellcolor[HTML]{FFFFFF}\textit{SoRFT-Qwen-32B}}                & \multicolumn{1}{c|}{\cellcolor[HTML]{FFFFFF}\textit{RL}}                         & \multicolumn{1}{c|}{\cellcolor[HTML]{FFFFFF}\textit{Pipeline}}                    & \multicolumn{1}{c|}{\cellcolor[HTML]{FFFFFF}\textcolor{orange}{30.8\%}}                                                 & \textcolor{orange}{24.0\%}                                    \\ \hline
\rowcolor[HTML]{FFFFFF} 
\multicolumn{1}{|c|}{\cellcolor[HTML]{FFFFFF}\textit{Lingma-SWE-GPT}~\cite{SWE-GPT}}                          & \multicolumn{1}{c|}{\cellcolor[HTML]{FFFFFF}\textit{Lingma-SWE-GPT-72B}}             & \multicolumn{1}{c|}{\cellcolor[HTML]{FFFFFF}\textit{SFT}}                        & \multicolumn{1}{c|}{\cellcolor[HTML]{FFFFFF}\textit{Agent}}                       & \multicolumn{1}{c|}{\cellcolor[HTML]{FFFFFF}30.2\%}                                                 & 22.0\%                                    \\ \hline
\rowcolor[HTML]{FFFFFF} 
\multicolumn{1}{|c|}{\cellcolor[HTML]{FFFFFF}\textit{SWE-GPT}~\cite{SWE-GPT}}                          & \multicolumn{1}{c|}{\cellcolor[HTML]{FFFFFF}\textit{SWE-Gym-Qwen-32B}}              & \multicolumn{1}{c|}{\cellcolor[HTML]{FFFFFF}\textit{SFT}}                        & \multicolumn{1}{c|}{\cellcolor[HTML]{FFFFFF}\textit{Agent}}                       & \multicolumn{1}{c|}{\cellcolor[HTML]{FFFFFF}20.6\%}                                                 & 15.3\%                                    \\ \hline
\rowcolor[HTML]{FFFFFF} 
\multicolumn{1}{|c|}{\cellcolor[HTML]{FFFFFF}\textit{SWE-Fixer}~\cite{SWE-Fixer}}                        & \multicolumn{1}{c|}{\cellcolor[HTML]{FFFFFF}\textit{SWE-Fixer-Qwen-72B}}            & \multicolumn{1}{c|}{\cellcolor[HTML]{FFFFFF}\textit{SFT}}                        & \multicolumn{1}{c|}{\cellcolor[HTML]{FFFFFF}\textit{Pipeline}}                    & \multicolumn{1}{c|}{\cellcolor[HTML]{FFFFFF}\textcolor{red}{32.8\%}}                                                 & \textcolor{red}{24.7\%}                                    \\ \hline

\rowcolor[HTML]{FFFFFF} 
\multicolumn{1}{|c|}{\cellcolor[HTML]{FFFFFF}\textit{Agentless}~\cite{Agentless}}                        & \multicolumn{1}{c|}{\cellcolor[HTML]{FFFFFF}\textit{Deepseek-V3-671B}}            & \multicolumn{1}{c|}{\cellcolor[HTML]{FFFFFF}\textit{SFT}}                        & \multicolumn{1}{c|}{\cellcolor[HTML]{FFFFFF}\textit{Pipeline}}                    & \multicolumn{1}{c|}{\cellcolor[HTML]{FFFFFF}42.0\%}                                                 & -                                    \\ \hline

\rowcolor[HTML]{FFFFFF} 
\multicolumn{1}{|c|}{\cellcolor[HTML]{FFFFFF}\textit{Agentless}~\cite{Agentless}}                        & \multicolumn{1}{c|}{\cellcolor[HTML]{FFFFFF}\textit{Deepseek-R1-671B}}            & \multicolumn{1}{c|}{\cellcolor[HTML]{FFFFFF}\textit{RL}}                        & \multicolumn{1}{c|}{\cellcolor[HTML]{FFFFFF}\textit{Pipeline}}                    & \multicolumn{1}{c|}{\cellcolor[HTML]{FFFFFF}49.2\%}                                                 & -                                    \\ \hline \hline

\rowcolor[HTML]{EFEFEF} 
\multicolumn{6}{|c|}{\cellcolor[HTML]{EFEFEF}\textit{\textbf{Fine-tuned Models based on our CoT Dataset}}}                                                                                                                                                                                                                                                                                                                                                                                    \\ \hline
\rowcolor[HTML]{FFFFFF} 
\multicolumn{1}{|c|}{\cellcolor[HTML]{FFFFFF}\textit{Agentless}~\cite{Agentless}}                        & \multicolumn{1}{c|}{\cellcolor[HTML]{FFFFFF}\textcolor{black}{\textbf{\textit{Fine-tuned Qwen2.5-7B-Instruct}}}}                    & \multicolumn{1}{c|}{\cellcolor[HTML]{FFFFFF}\textit{SFT}}                        & \multicolumn{1}{c|}{\cellcolor[HTML]{FFFFFF}\textit{Pipeline}}                    & \multicolumn{1}{c|}{\cellcolor[HTML]{FFFFFF}\textcolor{blue}{\textbf{22.6\%}}}                                                    & \textcolor{blue}{\textbf{16.3\%}}                                        \\ \hline
\rowcolor[HTML]{FFFFFF} 
\multicolumn{1}{|c|}{\cellcolor[HTML]{FFFFFF}\textit{Agentless}~\cite{Agentless}}                        & \multicolumn{1}{c|}{\cellcolor[HTML]{FFFFFF}\textcolor{black}{\textbf{\textit{Fine-tuned Qwen2.5-32B-Instruct}}}}                   & \multicolumn{1}{c|}{\cellcolor[HTML]{FFFFFF}\textit{SFT}}                        & \multicolumn{1}{c|}{\cellcolor[HTML]{FFFFFF}\textit{Pipeline}}                    & \multicolumn{1}{c|}{\cellcolor[HTML]{FFFFFF}\textbf{\textcolor{orange}{32.4\%}}}                                                    & \textcolor{orange}{\textbf{25.7\%}}                                        \\ \hline
\rowcolor[HTML]{FFFFFF} 
\multicolumn{1}{|c|}{\cellcolor[HTML]{FFFFFF}\textit{Agentless}~\cite{Agentless}}                        & \multicolumn{1}{c|}{\cellcolor[HTML]{FFFFFF}\textcolor{black}{\textbf{\textit{Fine-tuned Qwen2.5-72B-Instruct}}}}                 & \multicolumn{1}{c|}{\cellcolor[HTML]{FFFFFF}\textit{SFT}}                        & \multicolumn{1}{c|}{\cellcolor[HTML]{FFFFFF}\textit{Pipeline}}                    & \multicolumn{1}{c|}{\cellcolor[HTML]{FFFFFF}\textcolor{red}{\textbf{35.0\%}}}                                                    & \textcolor{red}{\textbf{28.3\%}}                                        \\ \hline
\end{tabular}
\begin{tablenotes}   
    \footnotesize            
    \item[1] '-' means the corresponding technique did not report \emph{\textbf{\%Resolved}} results in its paper.  '\emph{\text{N/A}}' denotes the unknown training technique.
   \item[2] \textbf{\emph{SFT}} denotes supervised fine-tuning; \textbf{\emph{RL}} denotes reinforcement learning.

   \item[3] \textcolor{blue}{\emph{Baselines' \textbf{\%Resolved}} values are from the technique's original paper.}
\end{tablenotes}

\egroup
\vspace{-4mm}
\label{RQ1}
\end{table}

\begin{figure*}[]
	\centering
    \vspace{-8mm}
\includegraphics[width=0.85\textwidth]{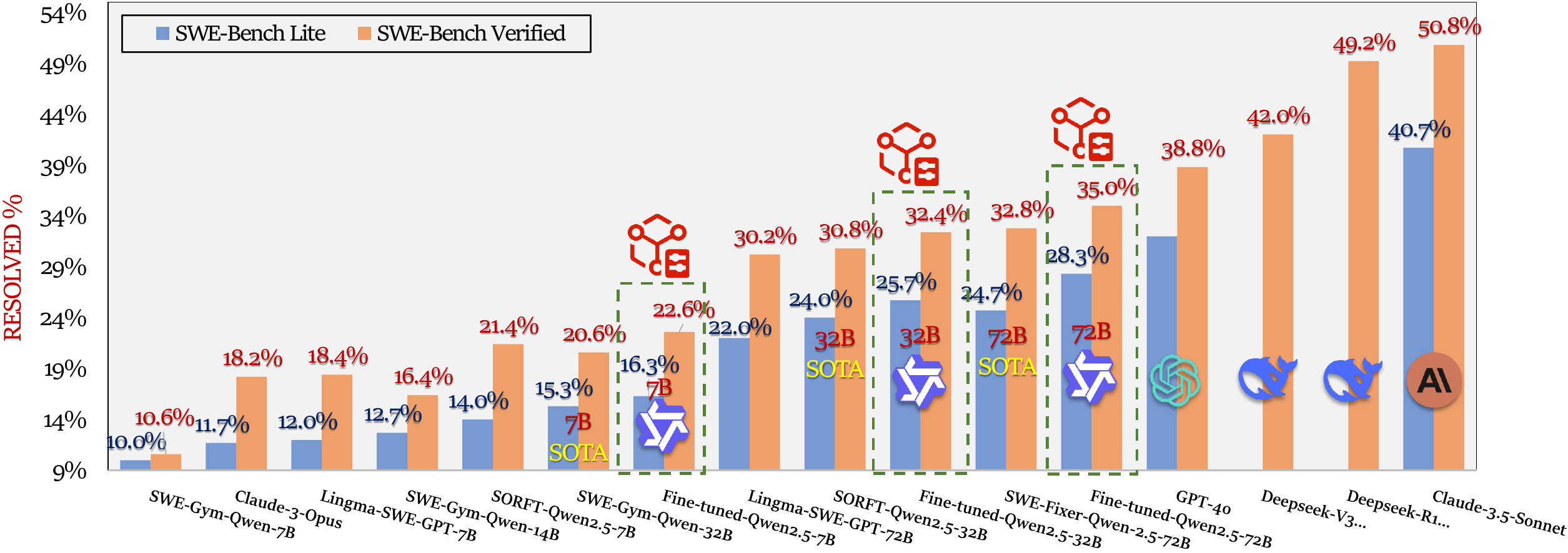}
	\vspace{-2mm}
	\caption{Comparison of issue resolution rates across baselines}
	\label{plot}
	\vspace{-6mm}
\end{figure*}

\noindent\textbf{\emph{Results.}} Table~\ref{RQ1} and Figure~\ref{plot} present the comparative performance of our fine-tuned model against existing approaches on both \emph{SWE-Bench Lite} and \emph{SWE-Bench Verified} benchmarks. The results are categorized into two distinct groups: proprietary model-based solutions and open-source model approaches. We observe that the highest resolution rates are predominantly achieved by closed-source model-based approaches. For instance, the \emph{OpenHands} method, which utilizes the \emph{Claude-3.5-Sonnet} model, achieves resolution rates of \textcolor{blue}{41.7}\% on \emph{SWE-Bench Lite} and \textcolor{blue}{53}\% on \emph{SWE-Bench Verified}.

\textbf{\emph{Comparison with Proprietary Model Frameworks.}} Notably, our fine-tuned 72B model outperforms several proprietary approaches on both \emph{SWE-Bench Lite} and \emph{Verified} benchmarks, including methods based on \emph{GPT-4}, \emph{GPT-4o}, and \emph{Claude-3-Opus}. For instance, the \textsc{SWE-agent} (using \emph{Claude-3-Opus}) achieves \textcolor{blue}{33.60}\% (\emph{Verified}) and \textcolor{blue}{23.00}\% (\emph{Lite}) resolution rates. In contrast, our approach yields higher performance: The fine-tuned 72B model achieves \textcolor{blue}{35.0}\% (\emph{Verified}) and \textcolor{blue}{28.3}\% (\emph{Lite}). Even the 32B variant reaches \textcolor{blue}{32.4}\% (\emph{Verified}) and \textcolor{blue}{25.7}\% (\emph{Lite}), demonstrating competitive efficiency.

\textbf{\emph{Comparison with Open-Source Model Frameworks.}} \emph{Lingma SWE-GPT} (based on \emph{Qwen2.5-72B-Instruct}) achieves resolution rates of \textcolor{blue}{22.0}\% (\emph{Lite}) and \textcolor{blue}{30.2}\% (\emph{Verified}). In comparison, our fine-tuned \emph{Qwen2.5-72B-Instruct} improves performance by \textcolor{blue}{\(\Uparrow\)}\textcolor{blue}{6.3}\% (\emph{Lite}) and \textcolor{blue}{\(\Uparrow\)}\textcolor{blue}{4.8}\% (\emph{Verified}). For the 32B model tier, existing approaches like \textsc{SORFT}~\cite{Sorft} (\emph{Qwen2.5-Coder-32B-Instruct}) report \textcolor{blue}{30.8}\% (\emph{Lite}) and \textcolor{blue}{24.0}\% (\emph{Verified}), whereas our fine-tuned variant boosts resolution rates by \textcolor{blue}{\(\Uparrow\)}\textcolor{blue}{1.6}\%  and \textcolor{blue}{\(\Uparrow\)}\textcolor{blue}{1.7}\%, respectively.
For models under 10B parameters, SOTA approach \textsc{SORFT}
~\cite{Sorft} achieves maximum resolution rates of \textcolor{blue}{21.4}\% (\emph{Lite}) and \textcolor{blue}{14.0}\% (\emph{Verified}). Our fine-tuned \emph{Qwen2.5-Coder-7B-Instruct} attains higher performance at \textcolor{blue}{\(\Uparrow\)}\textcolor{blue}{1.2}\% and \textcolor{blue}{\(\Uparrow\)}\textcolor{blue}{2.3}\%, respectively.
The results demonstrate that our fine-tuned models achieve significant improvements in issue resolution capability, establishing new SOTA performance within their respective parameter scales. 

\begin{figure}[]
	\centering
	\vspace{-2mm}
\includegraphics[width=0.40\textwidth]{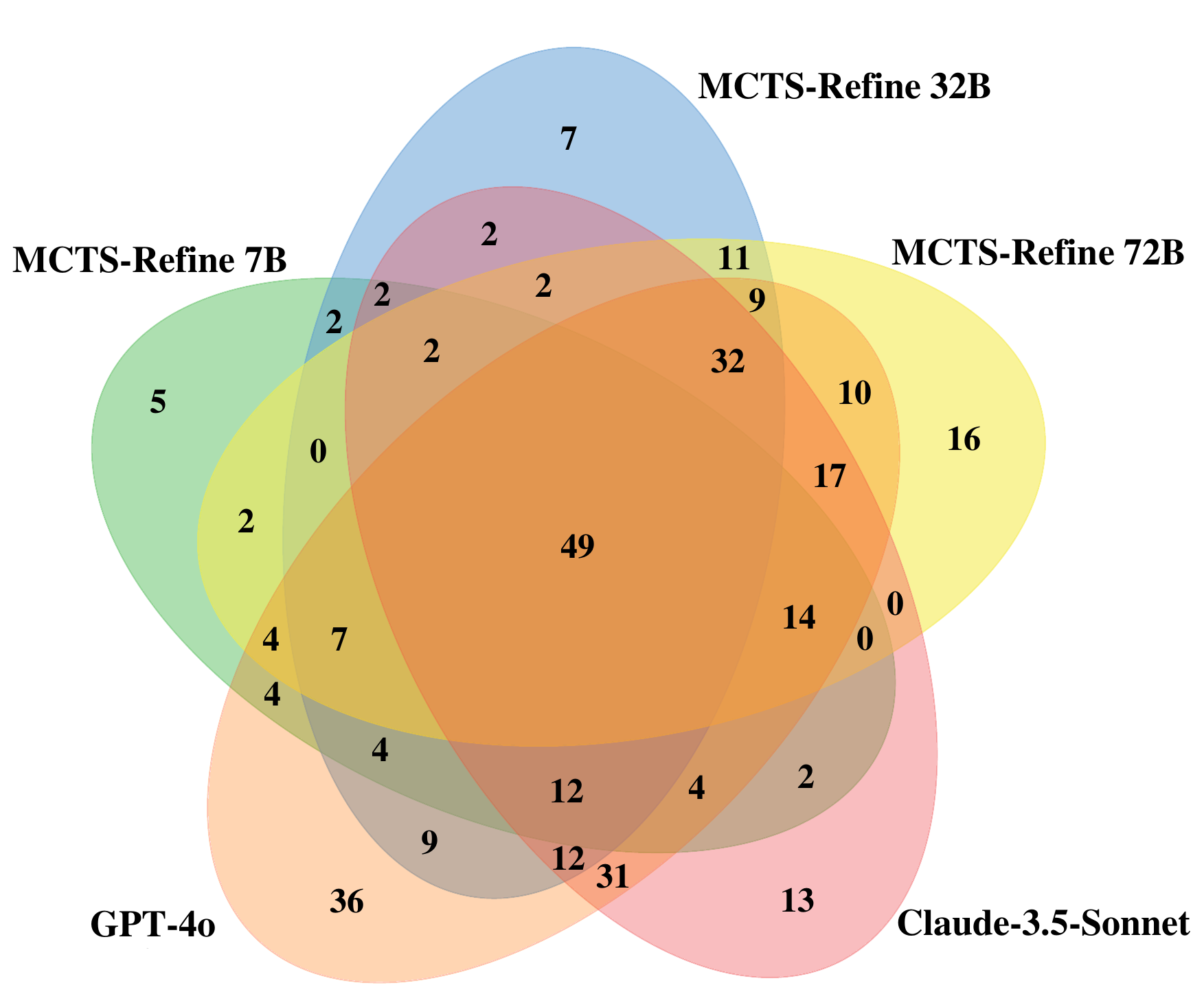}
	\vspace{-3mm}
	\caption{A Venn diagram examining the overlap of solved instances among different models on the \emph{SWE-bench Verified} benchmark}
	\label{venn}
	\vspace{-2mm}
\end{figure}

As shown in Figure~\ref{venn}, we conducted a Venn diagram analysis to examine the overlap of successfully solved instances among different models on the \emph{SWE-bench Verified} benchmark. The results reveal that our fine-tuned 7B, 32B, and 72B models uniquely resolved \textcolor{blue}{5}, \textcolor{blue}{7}, and \textcolor{blue}{16} issues respectively - cases that other models failed to address. Additionally, each model shares a substantial number of commonly solved instances with other models. These findings demonstrate that our approach not only achieves superior performance in terms of resolution quantity, but also exhibits unique issue-solving capabilities in terms of coverage scope.

Due to computational constraints, we adopted LoRA~\cite{hu2022lora} for fine-tuning \emph{Qwen2.5-72B-Instruct}, which achieved SOTA issue resolution rates at its parameter scale. However, empirical evidence shows that full-parameter fine-tuning consistently surpasses LoRA's performance given equivalent model capacity and training data~\cite{sun2023comparative}. Our analysis suggest that replacing LoRA with full-parameter fine-tuning on our CoT dataset could further improve the model's issue resolution capability.

\definecolor{mybg}{RGB}{204, 255, 204} 
\tcbset{
	colback=mybg,
        colframe=black,
	notitle, 
	width={\linewidth},
	top=0pt,
	left=1pt,
	right=1pt,
	bottom=0pt,
	toprule=0.5pt,
	titlerule=1pt,
	bottomrule=0.5pt,
	leftrule=0.5pt,
	rightrule=0.5pt,
	after skip=6pt,}
\noindent\begin{tcolorbox}
	\small
	\noindent\emph{\textbf{Conclusion:} Our \textbf{7B}, \textbf{32B}, and \textbf{72B} models, fine-tuned using MCoT data, demonstrate significant improvements in issue resolution, achieving state-of-the-art resolution rates for their respective parameter scales. On SWE-bench Verified, they achieve \textbf{\textcolor{blue}{16.3}\%}, \textbf{\textcolor{blue}{25.7}\%}, and \textbf{\textcolor{blue}{28.3}\%} resolution rates, while on SWE-bench Lite, they reach \textbf{\textcolor{blue}{22.6}\%}, \textbf{\textcolor{blue}{32.4}\%}, and \textbf{\textcolor{blue}{35.0}\%}.} 
\end{tcolorbox}

\subsection{RQ2: Effectiveness of File Location and Fault Localization}
\noindent\textbf{\emph{Methodology.}} 
Precise file and fault localization is fundamental to successful issue resolution, enabling both automated patch generation and developer debugging. To rigorously evaluate localization performance of our fine-tuned models, we leverage the \emph{SWE-Bench-Verified} dataset, comparing model-predicted patch locations against ground-truth patches across three granularity levels:

\begin{itemize}[leftmargin=*, topsep=1pt, itemsep=1pt]

\item \emph{\textbf{File Localization:}} A prediction is considered correct if the model’s suggested file list includes the file actually modified by developers.

\item \emph{\textbf{Function Localization:}} After integrating the patch into the codebase, we use Abstract Syntax Tree (AST) analysis to identify the modified functions or classes. A prediction is successful if the model’s output contains the exact function/class altered in the ground-truth patch.

\item \emph{\textbf{Line Localization:}} Following \emph{Lingma-SWE-GPT}~\cite{SWE-GPT}’s approach, we extract the patched line along with its three surrounding lines as the localization scope. A match is confirmed if this range includes the actual modified lines.

\end{itemize}


We conducted comprehensive evaluations of the fine-tuned \emph{Qwen2.5-Coder-7B-Instruct}, \emph{Qwen2.5-Coder-32B-Instruct}, and \emph{Qwen2.5-Coder-72B-Instruct} models, benchmarking their performance against open source approaches (including \emph{Lingma-SWE-GPT}~\cite{SWE-GPT}) and proprietary models (such as \emph{GPT-4o} and \emph{Claude 3.5 Sonnet}).
%
\begin{table}[]
\scriptsize
\centering
\setlength\tabcolsep{5pt}
\def\arraystretch{1}
\caption{Effectiveness of \emph{File Location} and \emph{Fault Localization}}
\vspace{-2mm}
 \bgroup
\begin{tabular}{|c|c|c|c|c|}
\hline
\rowcolor[HTML]{343434} 
{\color[HTML]{FFFFFF} \textit{\textbf{LLM}}}    & {\color[HTML]{FFFFFF} \textit{\textbf{Size}}} & {\color[HTML]{FFFFFF} \textit{\textbf{FileHit}}} & {\color[HTML]{FFFFFF} \textit{\textbf{FunHit}}} & {\color[HTML]{FFFFFF} \textit{\textbf{LineHit}}} \\ \hline \hline
\rowcolor[HTML]{FFFFFF} 
\textit{Qwen2.5-Coder-Instruct}                 & 7B                                            & 19.2\%                                           & 13.9\%                                          & 13.2\%                                           \\ \hline
\rowcolor[HTML]{FFFFFF} 
\textit{Lingma SWE-GPT}                         & 7B                                            & 58.8\%                                           & 42.2\%                                          & 39.1\%                                           \\ \hline
\rowcolor[HTML]{343434} 
{\color[HTML]{FFFFFF} \textit{Fine-tuned Qwen2.5-7B-Instruct}}  & {\color[HTML]{FFFFFF} 7B}                     & {\color[HTML]{FFFFFF} \textcolor{yellow}{64.4\%}}                         & {\color[HTML]{FFFFFF} \textcolor{yellow}{47.2\%}}                        & {\color[HTML]{FFFFFF} \textcolor{yellow}{41.8\%}}                         \\ \hline \hline
\rowcolor[HTML]{FFFFFF} 
\textit{Qwen2.5-Coder-Instruct}                 & 32B                                           & 33.2\%                                                & 28.8\%                                               & 25.4\%                                                \\ \hline
\rowcolor[HTML]{FFFFFF} 
\rowcolor[HTML]{343434} {\color[HTML]{FFFFFF} \textit{Fine-tuned Qwen2.5-32B-Instruct}}                        & {\color[HTML]{FFFFFF} 32B}                                           & {\color[HTML]{FFFFFF} \textcolor{green}{72.0\%} }                                               & {\color[HTML]{FFFFFF} \textcolor{green}{59.2\%}}                                              & {\color[HTML]{FFFFFF} \textcolor{green}{49.4\%} }                                               \\ \hline \hline
\rowcolor[HTML]{FFFFFF} 
\textit{Qwen2.5-Instruct}                       & 72B                                           & 61.3\%                                           & 45.7\%                                          & 42.8\%                                           \\ \hline
\rowcolor[HTML]{FFFFFF} 
\textit{Lingma SWE-GPT}                         & 72B                                           & 80.9\%                                           & 66.2\%                                          & 61.6\%                                           \\ \hline
\rowcolor[HTML]{343434} 
{\color[HTML]{FFFFFF} \textit{Fine-tuned Qwen2.5-72B-Instruct}} & {\color[HTML]{FFFFFF} 72B}                    & {\color[HTML]{FFFFFF} \textcolor{pink}{81.8\%}}                         & {\color[HTML]{FFFFFF} \textcolor{pink}{68.0\%}}                        & {\color[HTML]{FFFFFF} \textcolor{pink}{63.0\%}}                         \\ \hline \hline
\textit{GPT-4o}                                 & Unknown                                       & 72.5\%                                           & 55.7\%                                          & 52.2\%                                           \\ \hline
\textit{Claude-3.5-Sonnet}                      & Unknown                                       & 74.3\%                                           & 58.1\%                                          & 55.0\%                                           \\ \hline

\end{tabular}
\begin{tablenotes}   
    \footnotesize      
   \item[1] \textcolor{blue}{Due to computational constraints, our evaluation focuses on comparing localization results (at file/function/line levels) with established baselines from prior work~\cite{SWE-GPT}. 
   \emph{Baselines' \textbf{\%FileHit, \%FunHit, \%LineHit}} values are from \textsc{Lingma-SWE-GPT} study~\cite{SWE-GPT}.}
\end{tablenotes}
\egroup
\vspace{-6mm}
\label{RQ2}
\end{table} 

\noindent\textbf{\emph{Results.}} The results shown in Table~\ref{RQ2} demonstrate that the fine-tuned \emph{Qwen2.5-Coder} models achieve optimal performance across all granularity levels of localization tasks. Particularly noteworthy is the 72B model, which attains localization precision of \textcolor{blue}{81.8}\%, \textcolor{blue}{68.0}\%, and \textcolor{blue}{63.0}\% at the file, method, and line levels respectively - outperforming existing open-source issue resolution approaches like \emph{Lingma-SWE-GPT}.

At comparable parameter scales, both the fine-tuned 32B and 7B models consistently surpass other open-source models across all granularities. While proprietary models (\emph{GPT-4o} and \emph{Claude 3.5 Sonnet}) still maintain a performance advantage, the fine-tuned 72B model has achieved comparable or even superior localization accuracy to some proprietary models, further validating the effectiveness of both the \textsc{MCTS-Refine} framework and our fine-tuning strategy.

The fine-tuned models demonstrate the following localization capabilities across different granularities:

\begin{itemize}[leftmargin=*, topsep=1pt, itemsep=1pt]

\item \emph{\textbf{File Localization:}} The fine-tuned models exhibit superior precision in identifying issue-relevant files, particularly within complex software repositories, with the 72B model achieving a file localization precision of \textcolor{blue}{81.8}\%. This performance improvement is primarily attributed to the rigorous rejection sampling strategy employed by the \textsc{MCTS-Refine} framework during CoT data generation. 

\item \emph{\textbf{Function Localization:}} Our fine-tuned 72B model reaches \textcolor{blue}{68.0}\% accuracy in method localization, proving its enhanced capability in comprehending code semantics and structure.

\item \emph{\textbf{Line Localization:}} For line-level localization tasks, the fine-tuned 72B model achieves a precision of \textcolor{blue}{63.0}\%, outperforming \textsc{Lingma-SWE-GPT}~\cite{SWE-GPT}'s \textcolor{blue}{61.6}\%. This demonstrates the model's enhanced precision in identifying specific code modification locations, providing a reliable foundation for subsequent patch generation.

\end{itemize}

\noindent\begin{tcolorbox}
	\small
	\noindent\emph{\textbf{Conclusion:}} Our fine-tuned models achieve optimal performance across all granularity levels of localization tasks. Particularly noteworthy is the 72B model, which attains localization precision of \textbf{\textcolor{blue}{81.8}\%}, \textbf{\textcolor{blue}{68.0}\%}, and \textbf{\textcolor{blue}{63.0}\%} at the file, method, and line levels respectively - outperforming the existing open-source issue resolution approaches.
\end{tcolorbox}

\subsection{RQ3: Effectiveness of Patch Generation}
\noindent\textbf{\emph{Methodology.}} To assess the improvement of fine-tuned models in the patch generation subtask, we conducted a comparative evaluation between the fine-tuned \emph{Qwen2.5-Coder} models (\emph{7B-Instruct}, \emph{32B-Instruct}, and \emph{72B-Instruct}) and their corresponding base models. The experiment was performed on the \emph{SWE-bench Verified} benchmark with a specific focus on \emph{\textbf{Patch Generation}} capability. To evaluate patch generation performance, \textbf{we provided models with precise issue locations and 20 surrounding lines of contextual code}, requiring them to produce patches in strict \textbf{\emph{search-replace format}} (where "\emph{search}" identifies the original problematic code and "\emph{replace}" shows the corrected version) - a prerequisite for valid diff-based integration into original software repositories. 

Our two-phase assessment \textbf{\emph{first verifies format compliance}} (ensuring patches can be properly diff-applied), then \textbf{\emph{measures repair effectiveness through \%Resolved}}, thereby evaluating both syntactic correctness and actual issue-patching capability.

\begin{table}[]
\scriptsize
\centering
\setlength\tabcolsep{5pt}
\def\arraystretch{1}
\vspace{-8mm}
\caption{Effectiveness of \emph{Patch Generation}}
\vspace{-2mm}
 \bgroup
\begin{tabular}{|c|c|c|}
\hline
\rowcolor[HTML]{343434} 
{\color[HTML]{FFFFFF} \textit{\textbf{Models}}}         & {\color[HTML]{FFFFFF} \textit{\textbf{\begin{tabular}[c]{@{}c@{}}Format Compliance \\ Rate (\%)\end{tabular}}}} & {\color[HTML]{FFFFFF} \textit{\textbf{\begin{tabular}[c]{@{}c@{}}Resolution Rate\\ (\% Resolved)\end{tabular}}}} \\ \hline \hline
\rowcolor[HTML]{FFFFFF} 
\textit{Qwen2.5-Coder-Instruct 7B}             & 13.0\%                                                                                                               & 3.8\%                                                                                                                \\ \hline
\rowcolor[HTML]{C0C0C0} 
{\color[HTML]{000000} \textit{\textbf{Fine-tuned Qwen2.5-7B-Instruct}}} & {\color[HTML]{000000} \textcolor{red}{\textbf{94.6\%}}}                                                                                        & {\color[HTML]{000000} \textcolor{red}{\textbf{28.6\%}}}                                                                                         \\ \hline
\rowcolor[HTML]{FFFFFF} 
\textit{Qwen2.5-Coder-Instruct 32B}            & 23.4\%                                                                                                               & 9.6\%                                                                                                                \\ \hline
\rowcolor[HTML]{C0C0C0} 
\textit{\textbf{Fine-tuned Qwen2.5-32B-Instruct}}                       & \textcolor{red}{\textbf{96.0\%}}                                                                                                               & \textcolor{red}{\textbf{38.0\%}}                                                                                                                \\ \hline
\rowcolor[HTML]{FFFFFF} 
\textit{Qwen2.5-Instruct 72B}                  & 37.0\%                                                                                                               & 14.4\%                                                                                                                \\ \hline
\rowcolor[HTML]{C0C0C0} 
\textit{\textbf{Fine-tuned Qwen2.5-72B-Instruct}}                       & \textcolor{red}{\textbf{96.8\%}}                                                                                                               & \textcolor{red}{\textbf{43.8\%}}                                                                                                                \\ \hline
\rowcolor[HTML]{FFFFFF} 
\textit{Deepseek-V3}                  & 98.0\%                                                                                                              & 47.8\%                                                                                                                \\ \hline
\rowcolor[HTML]{C0C0C0} 
\textit{\textbf{Deepseek-R1}}                       & 98.6\%                                                                                                               & 59.0\%                                                                                                                \\ \hline
\end{tabular}
\begin{tablenotes}   
    \footnotesize      
   \item[1] \textcolor{black}{Since the SOTA techniques \emph{SORFT-Qwen-7B} and \emph{SORFT-Qwen-32B} are not open-sourced, and agent-based frameworks (e.g., \emph{Lingma SWE-GPT}, \emph{SWE-Agent}) cannot isolate the patch generation phase, we excluded them from baselines in RQ3. Instead, we compared the patch generation capabilities of \emph{Qwen2.5-Coder} models before and after fine-tuning.}
\end{tablenotes}
\egroup
\vspace{-6mm}
\label{RQ3}
\end{table} 

\noindent\textbf{\emph{Results.}} As evidenced in Table~\ref{RQ3}, fine-tuning yields significant improvements in patch generation performance. The pre-fine-tuned \emph{Qwen 2.5} models demonstrated poor performance in generating syntactically correct code edits, with the 72B variant producing only \textcolor{blue}{37.0}\% properly formatted patches and achieving merely \textcolor{blue}{14.4}\% issue resolution rate. In contrast, the fine-tuned models exhibit remarkable capabilities in generating both syntactically and structurally correct patches while substantially improving resolution rates. Notably, the fine-tuned 72B model achieves \textcolor{blue}{96.8}\% format compliance rate and \textcolor{blue}{43.8}\% resolution rate - significantly outperforming its untuned counterpart's \textcolor{blue}{37.0}\% and \textcolor{blue}{14.4}\%, respectively. Impressively, the fine-tuned 7B and 32B models surpass even the base 72B model in both edit format correctness and resolution rates. These results confirm that the \textsc{MCTS-Refine}-generated CoT dataset not only enhances the models' understanding of code syntax and structure, but also effectively improves their practical patch generation capabilities.

\noindent\begin{tcolorbox}
	\small
	\noindent\emph{\textbf{Conclusion:}} The fine-tuned models exhibit remarkable capabilities in generating correct patches. Notably, the fine-tuned 72B model achieves \textbf{\textcolor{blue}{96.8}\%} format compliance rate and \textbf{\textcolor{blue}{43.8}\%} resolution rate, \textbf{comparable to the performance of \emph{Deepseek-v3}}.
\end{tcolorbox}
\section{Threats to Validity}

\emph{\textbf{Data Leakage:}} ``Data leakage'' occurs when evaluation data overlaps with a LLM's training dataset, potentially causing overfitting and biased performance metrics. Although \emph{SWE-bench} derives from GitHub—the same platform used to train the \emph{Qwen2.5} foundation models—recent study~\cite{threats} suggests minimal leakage risk in this benchmark. For rigorous evaluation, we systematically exclude all \emph{SWE-bench} repositories when collecting CoT data from GitHub issue-pull requests.

\emph{\textbf{Reproducibility:}} To ensure reproducibility of LLM-generated content, we used the fixed open-source \emph{DeepSeek-V3-0324} model for CoT data generation in \textsc{MCTS-Refine} and publicly released the dataset. For \emph{SWE-bench} evaluation of our \textsc{Agentless} fine-tuned model, we maintained consistent parameters (\emph{temperature}=0) and used the official \emph{SWE-bench} Docker environment to guarantee reproducible results.

\vspace{-1mm}
\section{Related Works}

\emph{\textbf{{Training-Data Synthesis for LLMs.}}} To address the inefficiency of manual annotation, researchers have increasingly utilized large language models (LLMs) for scalable and high-quality data synthesis, capitalizing on their ability to efficiently generate large-scale datasets~\cite{toshniwal2024openmathinstruct,maini2024rephrasing,wang2024survey}. Recent work has focused on automating CoT data generation for issue-resolution tasks, aiming to enhance open-source models' reasoning via instruction tuning or reinforcement learning~\cite{SWE-GPT, SWE-Gym, Sorft, SWE-RL, SWE-Fixer}. \textsc{SWE-Gym}~\cite{SWE-Gym} synthesized \textsc{Openhands} Agent trajectories by prompting \emph{GPT-4o} and \emph{Claude-3.5-Sonnet} on open-source repositories, filtering outputs via unit tests before fine-tuning \emph{Qwen}. Similarly, \emph{Lingma-SWE GPT}~\cite{SWE-GPT} generated trajectories with \emph{GPT-4o} and filtered them by similarity, while \textsc{SWE-Fixer}~\cite{SWE-Fixer} and \textsc{SORFT}~\cite{Sorft} used single-turn Q\&A prompting (\emph{GPT-4o} and \emph{Claude-3.5-Sonnet}, respectively) to produce CoT data for edit generation and subtasks, applying overlap-based filtering. However, these approaches rely entirely on closed-source LLMs and suffer from critical limitations: their CoT synthesis lacks validation of intermediate reasoning steps and employs permissive rejection sampling (see discussions in Section~\ref{Limitations}). To overcome these limitations, we propose \textsc{MCTS-Refine}, a novel framework that replaces simplistic single-turn Q\&A prompting with a systematic, stepwise generation process, incorporating ground truth-aligned rejection sampling to rigorously validate each reasoning step and ensure robust, reliable reasoning-path synthesis.

\emph{\textbf{{LLMs for Repository Issue Resolution.}}} LLMs pretrained on massive code corpora have demonstrated exceptional capabilities in code-related tasks, from generation to repair~\cite{jiang2023impact,schafer2023empirical}, with recent frameworks like \emph{SWE-Bench}~\cite{SWE-Bench} extending these capabilities to repository-level problem solving through two dominant paradigms: {\textbf{(1) \emph{Agent-Based systems}} like \textsc{SWE-Agent}~\cite{SWE-Agent}, where LLMs autonomously interact with development environments via specialized interfaces (e.g., Agent-Computer Interfaces) to perform editing, navigation, and testing; and {\textbf{(2) \emph{Pipeline-Based Approaches:}} such as \textsc{RAG-SWE}~\cite{SWE-Bench} and \textsc{Agentless}~\cite{Agentless}, which employ structured workflows combining retrieval, context aggregation, and targeted editing. Agent-based systems enable broader applications but demand advanced LLMs and multi-turn computation, while pipeline methods boost efficiency through focused, staged task execution.
In this paper, we integrate our fine-tuned model into the \textsc{Agentless 1.0}~\cite{Agentless} framework, which is compatible with our CoT data, enabling end-to-end evaluation of \textsc{MCTS-Refine}'s \emph{Rejection Sampling} and \emph{Refinement} methodology.

\vspace{-2mm}
\section{Conclusion}

We proposed \textsc{MCTS-Refine}, a novel framework that enhances open-source LLMs' issue resolution capabilities through high-quality CoT data generation. By combining MCTS with a reflection mechanism and rigorous subtask validation, our approach achieves state-of-the-art performance on \emph{SWE-bench}. The released resources - including the \textsc{MCoT} dataset and fine-tuned models - provide valuable tools for advancing AI-powered software engineering. 
\section*{Acknowledgment}

This work was supported by the Tencent Rhino-Bird Basic Research Program (Project No. JR2024TEG004), the National Key Research and Development Program of China (Grant No. 2024YFF0908000) and the China Postdoctoral Science Foundation (Grant No. 2024M750375). 

\end{document}